\documentclass[twocolumns,a4paper]{aa}
\usepackage{graphicx}
\usepackage{txfonts}
\usepackage{color}
\usepackage{natbib}
\usepackage{hyperref}

\usepackage{amsmath}

\definecolor{blue}{RGB}{0,0,255}
\definecolor{red}{RGB}{255,0,0}
\definecolor{green}{RGB}{0,255,0}
\newcommand{\ro}[1]{\textcolor{red}{#1}}

\begin{document}

\title{Combining magneto-hydrostatic constraints with Stokes profile inversions.
  \\ II. Application to Hinode/SP observations}
\author{J.M.~Borrero\inst{1} \and A.~Pastor Yabar\inst{2} \and
  B.~Ruiz Cobo\inst{3,4}}
\institute{Leibniz-Institut f\"ur Sonnenphysik, Sch\"oneckstr. 6, D-79110, Freiburg, Germany
\and
Institute for Solar Physics, Department of Astronomy, Stockholm University, AlbaNova University 
Centre, 10691 Stockholm, Sweden
\and
Instituto de Astrof{\'\i}sica de Canarias, Avd. V{\'\i}a L\'actea s/n, E-38205, La Laguna, Spain
\and
Departamento de Astrof{\'\i}sica, Universidad de La Laguna, E-38205, La Laguna, Tenerife, Spain
}
\date{Recieved / Accepted}

\abstract{Inversion techniques applied to the radiative transfer equation for polarized light are
capable of inferring the physical parameters in the solar atmosphere (temperature $T$, magnetic field ${\bf B}$,
and line-of-sight velocity $v_{\rm los}$) from observations of the Stokes vector (i.e., spectropolarimetric observations) 
in spectral lines. Inferences are usually performed in the $(x,y,\tau_c)$ domain, where $\tau_c$ refers to the optical-depth scale. 
Generally, their determination in the $(x,y,z)$ volume is not possible due to the lack of a reliable estimation of the gas 
pressure, particularly in regions of the solar surface harboring strong magnetic fields.}
{We aim to develop a new inversion code capable of reliably inferring the physical parameters in the $(x,y,z)$ domain.}{We combine, in a self-consistent way, an inverse solver for the radiative 
transfer equation (Firtez-DZ) with a solver for the magneto-hydrostatic (MHS) equilibrium, which
derives realistic values of the gas pressure by taking the magnetic pressure and tension  into account.}
{We test the correct behavior of the newly developed code with spectropolarimetric observations of two 
sunspots recorded with the spectropolarimeter (SP) instrument on board the Hinode spacecraft, and we show how the physical
parameters are inferred in the $(x,y,z)$ domain, with the Wilson depression of the sunspots arising
as a natural consequence of the force balance. In particular, our approach significantly improves upon previous
determinations that were based on semiempirical models.}{Our results open the door for the possibility of
calculating reliable electric currents in three dimensions, ${\bf j}(x,y,z)$, in the solar photosphere.
Further consistency checks would include a comparison with other methods that have recently been proposed and which achieve similar goals.}

\titlerunning{MHS constraints in Stokes inversions. II. Application to Hinode/SP}
\authorrunning{Borrero et al.}
\maketitle

\def\kms{~km s$^{-1}$}
\def\deg{^{\circ}}
\def\df{{\rm d}}
\newcommand{\ve}[1]{{\rm\bf {#1}}}
\newcommand{\diff}{{\rm d}}
\newcommand{\Conv}{\mathop{\scalebox{1.5}{\raisebox{-0.2ex}{$\ast$}}}}%
\def\ex{{\bf e_x}}
\def\ez{{\bf e_z}}
\def\ey{{\bf e_y}}
\def\expr{{\bf e_x^\ensuremath{\prime}}}
\def\ezpr{{\bf e_z^\ensuremath{\prime}}}
\def\eypr{{\bf e_y^\ensuremath{\prime}}}
\def\xp{x^\ensuremath{\prime}}
\def\yp{y^\ensuremath{\prime}}
\def\zp{z^\ensuremath{\prime}}
\def\rp{r^\ensuremath{\prime}}
\def\xas{x^{\ast}\!}
\def\yas{y^{\ast}\!}
\def\zas{z^{\ast}\!}
\def\C{\mathcal{C}}

\section{Introduction}
\label{sec:introduction}

Inversion techniques applied to the radiative transfer equation for polarized light are arguably the best
tools at our disposal for inferring the physical properties (temperature $T$, magnetic field ${\bf B}$, and line-of-sight
velocity $v_{\rm los}$) of the solar atmosphere \citep{hector2001review,jc2003review,luis2006review,basilio2007review,jc2016review}. 
Because the natural scale to describe how photons propagate is the so-called optical depth ($\tau$), the physical properties are inferred in
the $(x,y,\tau_c)$, where $\tau_c$ refers to the continuum optical depth. Here "continuum" means any wavelength
where the absorption is only due to bound-free and free-free transitions \citep[][Sect.~4.4]{mihalas1970}.\\

In order to infer the physical parameters in the $(x,y,z)$ domain, additional constraints must be invoked. By far,
the most widely used has been hydrostatic equilibrium. However, this assumption is adequate only in 
regions where the magnetic field is force-free (i.e., Lorentz force $\propto {\bf j} \times {\bf B} = 0$) and the
plasma is stationary (i.e., no time dependence) and static (i.e., no velocities). In many regions of the solar atmosphere,
notably in sunspots, the force-free assumption breaks down and a different method must therefore be employed.\\

The first authors that attempted a more realistic treatment were \cite{valentin1990zw,valentin1993zw,solanki1993zw}.
They all employed the theoretical model from \cite{maltby1977zw}, which considers an axially symmetric magnetic field
around the sunspot in order to account for the magnetic pressure and tension. This approach had been used until recently
\citep[see e.g.,][]{mathew2004zw}, until the pioneering work of \cite[][hereafter referred to as PUS2010]{puschmann2010pen}, 
who presented a new method based on the minimization of the Lorentz force and ${\bf \nabla} \cdot {\bf B}$ in order 
to transform the physical parameters from the $(x,y,\tau_c)$ domain into the $(x,y,z)$ domain. Despite its importance, 
PUS2010 suffers a couple of drawbacks. The first is that the minimization,
based on a genetic algorithm, is very slow due to the large number of free parameters and therefore can only deal 
with relatively small regions. More important, however, is the fact that the gas pressure is modified in the process of inferring the physical 
parameters in the $(x,y,z)$ domain, and therefore the physical parameters are not able to provide the best 
possible fit to the observed polarization signals.\\

The results from PUS2010 have sparked a new interest in developing an inversion code for the radiative transfer
equation that is capable of inferring the physical parameters in the solar atmosphere in the geometrical $(x,y,z)$ three-dimensional
domain. This has resulted in a number of new approaches, beginning with adapting the PUS2010 method to minimize only 
${\bf \nabla} \cdot {\bf B}$ but in a much larger area \citep{loeptien2018zw,loeptien2020zw}. Methods that rely on
magneto-hydrodynamic (MHD) simulations have also been developed, with some using these simulations as training sets for
artificial neural networks \citep{carroll2008invz} and convolutional neural networks \citep{andres2019invz}, while others 
employing them as a database of physical parameters capable of fitting the observed polarization signals \citep{tino2017invz}.\\

We have developed an alternative approach that is loosely based on PUS2010. In \cite{adur2019invz}, we presented 
an inversion code for the radiative transfer equation that works directly in the $(x,y,z)$ domain, and we showed that the 
reliability of inferences in the $z$-scale depend upon the realism of the gas pressure ($P_{\rm g}$). In \cite{borrero2019mhs}, 
we presented a method that is based on the magneto-hydrostatic (MHS) equilibrium instead of hydrostatic equilibrium and can be used to 
infer very realistic values of $P_{\rm g}$. In this article, we come full circle and demonstrate how the approaches presented 
in the previous two papers can be combined to determine accurate physical parameters in the solar atmosphere in the $(x,y,z)$ 
domain, by applying our newly developed methods to spectropolarimetric observations with high spatial and spectral resolution.

\begin{figure*}
\begin{center}
\begin{tabular}{cc}
\includegraphics[width=8cm]{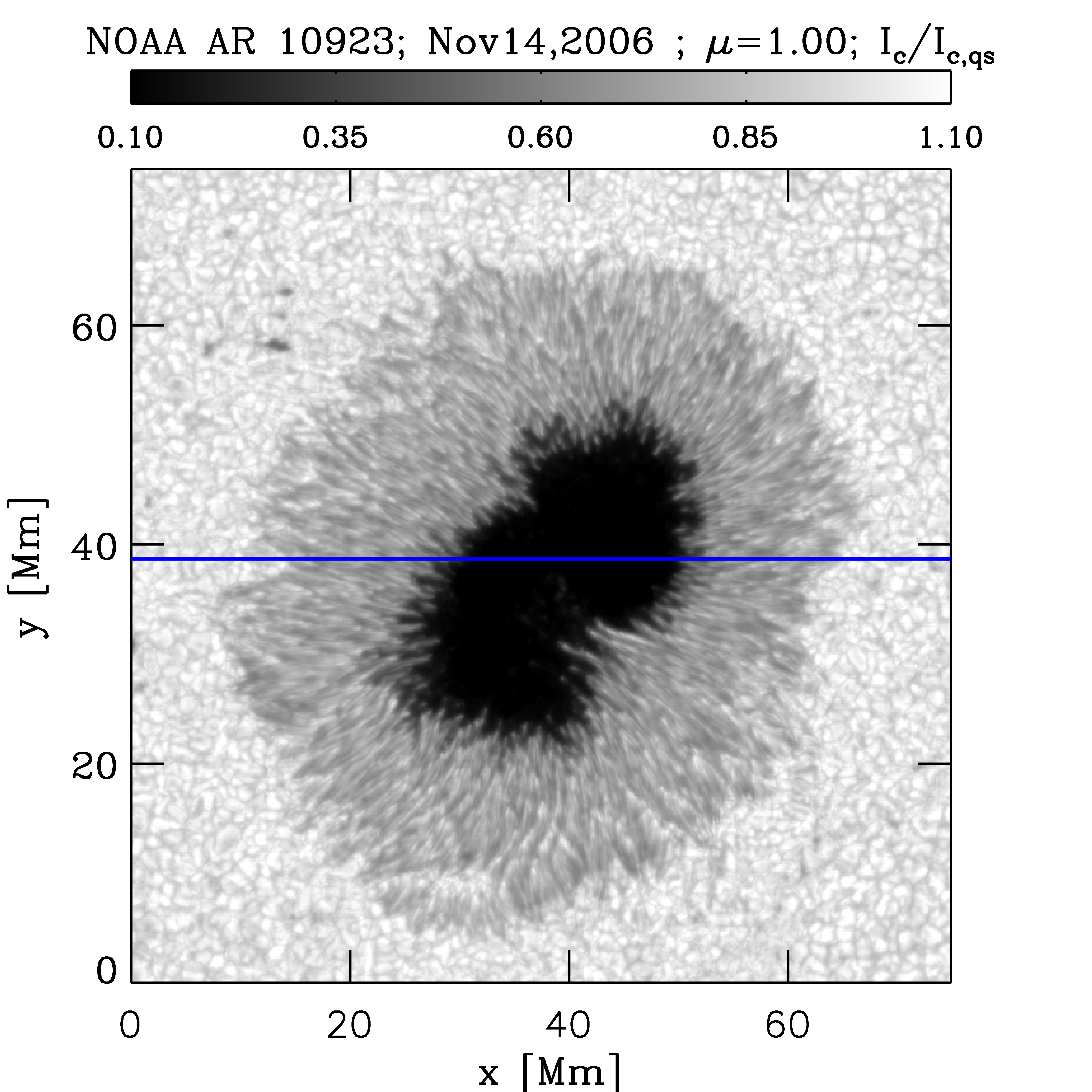} &
\includegraphics[width=8cm]{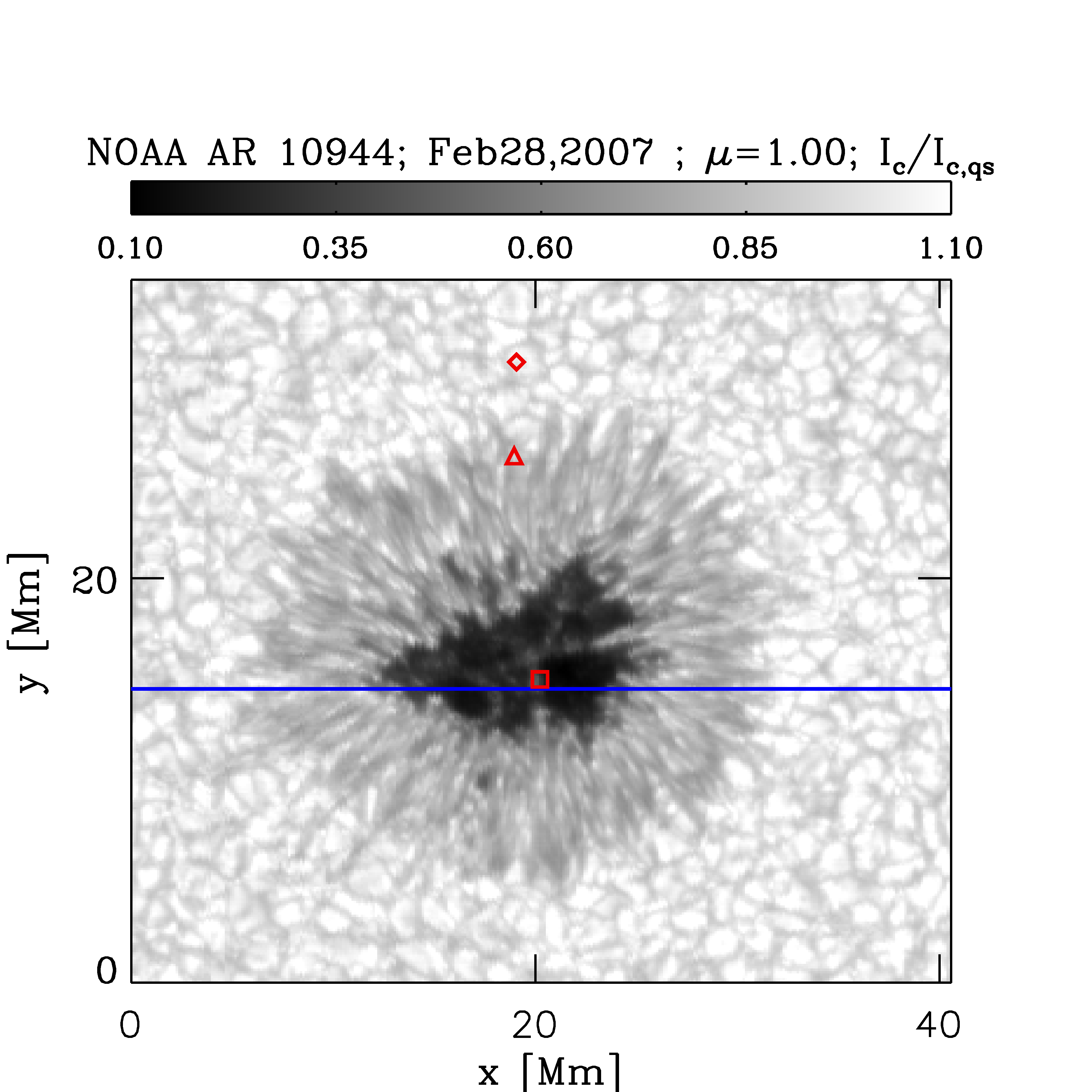}
\end{tabular}
\caption{Maps of the normalized continuum intensity, $I_{\rm c}/I_{\rm c,qs}$, for the two sunspots
analyzed in this work. {\it Left}: NOAA AR 10923 observed on November 14, 2006. {\it Right}: NOAA
AR 10944 observed on February 28, 2007. At the time of the observations, both sunspots were
located at disk center. Regions marked with red symbols and solid blue lines will be studied in more
detail later on.\label{fig:icmaps}}
\end{center}
\end{figure*}

\section{Hinode/SP observations}
\label{sec:observations}

The observations employed in this work correspond to spectropolarimetric observations
(i.e., Stokes vector ${\varmathbb I}^{\rm obs}$) of two neutral iron (Fe {\sc I}) spectral lines 
at 630 nm. The Stokes vector possesses four components, 
${\varmathbb I}=(I,Q,U,V)$, where $I$ refers to the total intensity, $Q$ and $U$ to the linear 
polarization, and $V$ to the circular polarization \citep[see Sect.~3.3 in][]{deltoro2003book}.

The observations were carried out with the spectropolarimeter \citep[SP;][]{lites2001hinode,ichimoto2007hinode}
attached to the Solar Optical Telescope \citep[SOT][]{suematsu2008hinode,tsuneta2008hinode,shimizu2008hinode}
on board the Japanese satellite Hinode \citep{kosugi2007hinode}. The spectral region containing the
two aforementioned Fe {\sc I} lines was measured across $\Lambda=112$ wavelength points with a wavelength sampling of about 
21.5 m{\AA}. The atomic parameters for these spectral lines can be found in \cite{borrero2014milne} (see their Table 1).
The SP is a slit-spectrograph where a given region is scanned spatially. For each slit
position, the light is integrated for a total of 4.8 seconds, yielding a noise level of $\sigma = 10^{-3}$
in units of the quiet-Sun continuum intensity. The spatial sampling along the slit and perpendicular to it
is about 0.16 arcsec (i.e., $\df x = \df y = 120$~km at disk center).\\

In this work, we analyze spectropolarimetric data from two different sunspots: NOAA AR 10923 and NOAA AR 10944. 
Both spots were observed very close to disk center $\mu \approx 1.0$, on November 14, 2006 (at around 7:15 UT)
and February 28, 2007 (at around 11:50 UT), respectively. Maps of the continuum intensity $I_{\rm c}$,
normalized to the quiet-Sun continuum intensity $I_{\rm c,qs}$, can be seen in Fig.~\ref{fig:icmaps}
for AR 10923 (left) and AR 10944 (right). The analyzed maps possess the following horizontal 
dimensions (in pixels): $L=645$, $M=640$ and $L=350$, $M=300$, respectively.\\

\section{Methodology}
\label{sec:methodology}

\subsection{Stokes inversion with Firtez-DZ}
\label{sec:firtez}

\begin{figure*}
\begin{center}
\includegraphics[width=18cm]{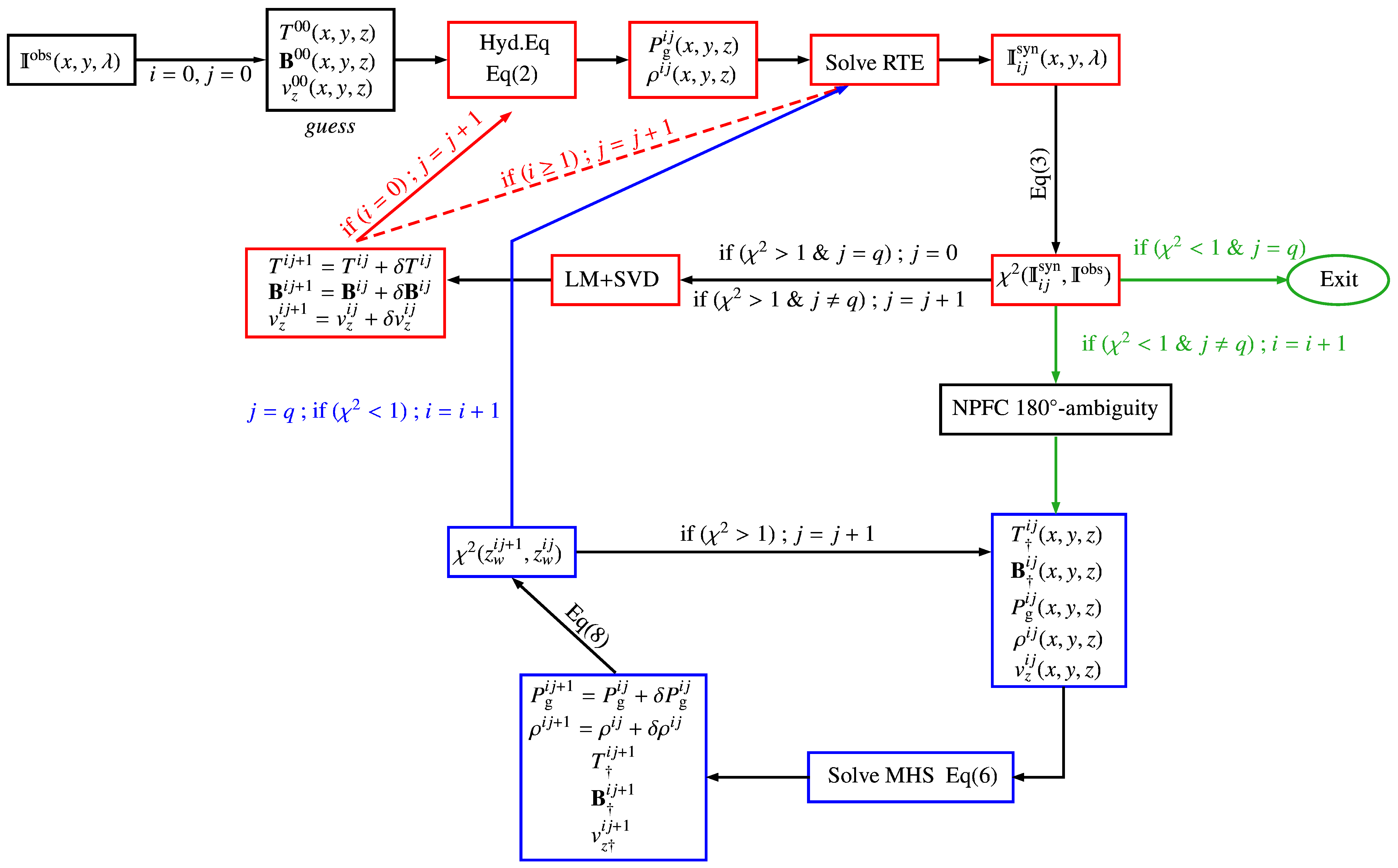}
\caption{Flow chart indicating the inversion process of the radiative transfer equation (RTE) for polarized
light (i.e., the Stokes inversion) combined with MHS constraints. The black squares denote the 
acquisition of the observed Stokes vector $\varmathbb{I}^{\rm obs}(x,y,\lambda)$ and the determination of an initial set
of physical parameters ($T^{00}(x,y,z)$, ${\bf B}^{00}(x,y,z)$) with which we can start the inversion (i.e., a guess).
The red squares indicate the inversion process as carried out by the Firtez-DZ code. This is described in detail
in Sect.~\ref{sec:firtez}. Blue squares correspond to the steps carried out by the MHS module (see Sect.~\ref{sec:mhs}
for details). Finally, green squares and arrows indicate locations where an interplay between Firtez-DZ and the MHS
module are needed in order to assess if convergence and exit conditions are achieved (see Sect.~\ref{sec:iter} for more information).
\label{fig:sketch}}
\end{center}
\end{figure*}

The Stokes inversion code employed in this work is Firtez-DZ \citep{adur2019invz}. A graphical sketch of how Firtez-DZ 
operates is presented in Fig.~\ref{fig:sketch} and is highlighted in red boxes. A more detailed description of this 
figure will be given throughout this section. Firtez-DZ needs guesses of 
the physical parameters $\C^{ij}$ in the solar atmosphere as inputs. We refer to these
physical parameters with the super-indexes $i,j$, where $i$-even indicates that we are
currently inside the Stokes inversion loop within Firtez-DZ, while $i$-odd implies that we are inside
the MHS module. Index $j$ stands for the iteration number within either Firtez-DZ or the MHS module 
and is reset to $j=0$ every time the Stokes inversion and MHS modules communicate with each other.\\

The aforementioned physical parameters $\C^{ij}$ stand for: temperature ($T^{ij}$), three components of the 
magnetic field ($B_x^{ij}$, $B_y^{ij}$, $B_z^{ij}$), and the line-of-sight component of the velocity ($v_{\rm los}^{ij}$), 
all as a function of the Cartesian \footnote{In this paper we will always assume that the observer's line-of-sight 
is parallel to the gravity direction $-z$ and therefore $v_{\rm los}=v_z$. This is possible because the selected
observations are very close to disk center ($\mu \approx 1$; see Sect.~\ref{sec:observations})} coordinates $(x,y,z)$. 
Besides these physical parameters, Firtez-DZ needs the density $\rho^{ij}$ and gas pressure $P_{\rm g}^{ij}$. The 
former can be obtained from the latter if the temperature is known by using the equation of state:

\begin{equation}
\rho^{ij} = \frac{u}{K_b}\frac{\mu^{ij} P_{\rm g}^{ij}}{T^{ij}} \;\;,
\label{eq:eos}
\end{equation}

\noindent where $u$ and $K_b$ refer to the atomic unit mass and the Boltzmann constant, respectively:
$u=1.6605\times 10^{-24}$ g and $K_b=1.3806\times 10^{-16}$ erg~K$^{-1}$. The mean molecular weight
$\mu$ is a function of $T$ and $P_{\rm g}$, and its determination involves the iterative computation
of the Saha ionization equation and the Boltzmann equation for the occupancy of the energy levels within 
an atom \citep{mihalas1970}.\\

The question that remains is how to determine the gas pressure $P_{\rm g}^{ij}$ at every $j$-step
during the inversion process. At $i=0$, the MHS module has not yet been employed, so we need to rely
on hydrostatic equilibrium approximation along the vertical direction:

\begin{equation}
\frac{\partial P_{\rm g}^{0j}}{\partial z} = -\rho^{0j} g \;\; \:\;,
\label{eq:hydeq}
\end{equation}

\noindent where $g=2.74$ cm~s$^{-2}$ is the Sun's gravitational acceleration. The gas pressure
is recalculated at every $j$-step during the Stokes inversion as long as $i=0$ (i.e., hydrostatic
equilibrium). This is indicated by the solid red arrow in Fig.~\ref{fig:sketch}. For $i \ge 1$, the MHS 
module (Sect.~\ref{sec:mhs}) already provides the gas pressure, and therefore we do not need to 
calculate it. Indeed, for $i \ge 1$, Firtez-DZ keeps $P_{\rm g}$ constant during the Stokes 
inversion (i.e., $j$-step; see dashed red arrow in Fig.~\ref{fig:sketch}).\\

With all these ingredients, Firtez-DZ solves the radiative transfer equation for polarized light in the $z$-scale 
\citep{egidio1985rte} under the assumption of local thermodynamic equilibrium and computes the polarized spectrum 
(i.e., Stokes vector ${\varmathbb I}_{ij}$) of atomic spectral lines in the Zeeman regime as a function of 
wavelength and horizontal grid position ($x$,$y$,$\lambda$). This Stokes vector is referred to as a "synthetic" Stokes vector and is denoted as ${\varmathbb I}^{\rm syn}_{ij}(x,y,\lambda)$. The four
components of the Stokes vector (see Sect.~\ref{sec:observations}) are generically referred to as $I_{s,ij}$
($I_{s=1}=I, I_{s=2}=Q, I_{s=3}=U, I_{s=4}=V$). The ${\varmathbb I}^{\rm syn}_{ij}(x,y,\lambda)$ is then compared to the 
observed Stokes vector ${\varmathbb I}^{\rm obs}(x,y,\lambda)$ via a $\chi^2$-merit function:

\begin{eqnarray}
\begin{split}
\chi^2(\varmathbb{I}^{\rm syn}_{ij},\varmathbb{I}^{\rm obs}) = \frac{1}{4ML\Lambda-F} & \sum\limits_{l=1}^{L}\sum\limits_{m=1}^{M}
\sum\limits_{k=1}^{\Lambda}\sum\limits_{s=1}^4 w_s^2 [I_s^{\rm obs}(x_l,y_m,\lambda_k)- \\ & I_{s,ij}^{\rm syn}(x_l,y_m,\lambda_k)]^2 
\;\;\textrm{with $i$ even},
\label{eq:chi2i}
\end{split}
\end{eqnarray}

\noindent where the sum runs for all grid points on the horizontal plane $(x,y)$ (indexes $l$ and $m$, respectively), for all observed 
wavelengths (index $k$) and for all four Stokes parameters (index $s$). In order to help the reader keep track of all indexes, a summary 
is provided in Table~\ref{table:index}. In Eq.~\ref{eq:chi2i}, $F$ stands for the total number of free parameters employed in the 
inversion (see Table~\ref{table:nodes}).  The $w_s$ factors in Eq.~\ref{eq:chi2i} are used as weights during the inversion of the 
radiative transfer equation \citep[see Eq.~35][]{jc2016review}, and $\chi^2$ is normalized such that a value of $\chi^2<1$ indicates a 
good fit between $\varmathbb{I}^{\rm obs}$ and $\varmathbb{I}^{\rm syn}_{ij}$. In this paper, the inversion is performed such that it gives three 
times more weight to the linear polarization profiles $Q$ and $U$ than to $I$: $w_2=w_3=3 w_1$, and two times more weight to the 
circular polarization $V$ than to $I$: $w_4=2 w_1$. The weight given to Stokes $I$ was taken as the inverse of the noise 
(see Sect.~\ref{sec:observations}): $w_1=1/\sigma$.\\

Analytical derivatives of $\chi^2$ with respect to the physical parameters\footnote{These derivatives are ultimately written
as a function of the derivatives of the Stokes vector with respect to the physical parameters: the so-called response functions
\citep[see Sect.~6 in][]{jc2016review}.} are calculated and fed into a Levenberg-Marquardt (LM) algorithm 
\citep[]{press1986num} that, along with the singular 
decomposition value (SVD) method \citep[][]{golub1965svd}, provides the new physical parameters in the solar 
atmosphere $\C^{ij+1}$ as a function of $(x,y,z); $ these new parameters produce a better match between the synthetic
and observed Stokes profiles: $\chi^2_{ij+1} < \chi^2_{ij}$. This process continues iteratively until 
the best possible match between the synthetic and observed Stokes vector is found (i.e., $\chi^2$-minimization).\\

We will now assume that the minimization is achieved after $j=p$ iterations of the Stokes inversion process ($i$-even),
thus proving the physical parameters in the solar atmosphere, $[\C^{ip}, P_{\rm g}^{ip}, \rho^{ip}]$, as a function of $(x,y,z)$. 
If $i=0$, Firtez-DZ provides only a "first estimation" of the physical parameters in the solar atmosphere as a function
of $(x,y,z)$ because, as discussed in \cite{adur2019invz}, their reliability in the $(x,y,z)$ domain depends upon the 
accuracy of the gas pressure $P_{\rm g}(x,y,z)$, whose inference is in turn hindered by the limitations of 
hydrostatic equilibrium employed at $i=0$. In order to improve the determination of $P_{\rm g}(x,y,z)$,
all physical parameters ($T^{ip}$, $P_{\rm g}^{ip}$, $\rho^{ip}$, $B_x^{ip}$, $B_y^{ip}$, $B_z^{ip}$, and $v_z^{ip}$) are 
then passed onto the disambiguation module (Sect.~\ref{sec:disamb}) and from there to the 
MHS module (Sect.~\ref{sec:mhs}). With this, we increase the $i$-index by one ($i$ is now odd), and,
since the MHS module has its own internal iteration that is independent from the Stokes inversion, we also 
reset the $j$-index to zero. This step is indicated by the green arrow in Fig.~\ref{fig:sketch}.\\

During the inversion process, the three-dimensional volume is discretized in $L$, $M$, and $N$ points along each
of the three Cartesian coordinates, $x$, $y$, and $z$, respectively. The grid sizes are denoted
as $\df x, \df y$, and $ \df z$. In all our inversions, we discretized the vertical direction with
$N=128$ grid points with a spacing of $\df z=12$ km. The number of grid points on the horizontal plane, 
$L$ and $M$, depends on the actual size of the observed sunspots (see Sect.~\ref{sec:observations}).
The horizontal spacing is always $\df x = \df y = 120$~km. A summary of these values is also included
in Table~\ref{table:index}.\\

\begin{table}
\caption{Summary of indexes employed in Sect.~\ref{sec:methodology}.\label{table:index}}
\begin{center}
\begin{tabular}{cccc}
index & phys.mag & ref. & step size \\
\hline
$i$-even, $j$-any & $\chi^2(\varmathbb{I}^{\rm syn}_{ij},\varmathbb{I}^{\rm obs})$ & Sect.~\ref{sec:firtez}; Eq.~\ref{eq:chi2i} & na\\
$i$-odd, $j$-any & $\chi^2(z_w^{ij+1},z_w^{ij})$ & Sect.~\ref{sec:mhs}; Eq.~\ref{eq:chi2z} & na\\
$k=1,...,\Lambda=112$ & $\varmathbb{I}$ & Sect.~\ref{sec:observations} & 21.5 m{\AA} \\
$s=1,...,4$ & $\varmathbb{I}$ & Sect.~\ref{sec:firtez} & na \\
$l=1,...,L$ & $x$ & Sect.~\ref{sec:observations} & 120 km \\
$m=1,...,M$ & $y$ & Sect.~\ref{sec:observations} & 120 km \\
$n=1,...,N=128$ & $z$ & Sect.~\ref{sec:firtez} & 12 km \\
\end{tabular}
\end{center}
\end{table}

We note that the inversion process performed by Firtez-DZ is done in such a way that the complexity
of the atmospheric model along the vertical $z$-direction increases slowly. This means that the number of
free parameters that are determined, at every $j$-step of the Stokes inversion process ($i$-even), also 
increases. More details can be found in \cite[][see Sect.~2.3]{adur2019invz}. The number of free parameters 
employed in this paper is indicated in Table~\ref{table:nodes}.\\

We slightly modified the original implementation of Firtez-DZ in order to avoid excessively modifying 
the temperature outside the "sensitivity region," which we denote as $[\tau_a,\tau_b]$ 
($\tau_a > \tau_b$; see also Appendix~\ref{app:sensi}). To do so, temperature perturbations 
$\delta T$, calculated with the LM algorithm, are forced to exponentially decay above $\tau_b$:

\begin{equation}
\delta T(z)=\delta T(z[\tau_{i}])\,exp\{-2(\log\tau_{b}-\log\tau_{i})\} \;\; 
\textrm{if} \;\; z <z(\tau_b)
.\end{equation}
Additionally, temperature perturbations for layers below $\tau_a$ are set to be equal to those at the 
sensitivity region limit, namely: $\delta T(z)=\delta T(z[\tau_{a}])$ if $z > z(\tau_a)$.

\begin{table}
\caption{Summary of free parameters in Firtez-DZ (Sect.~\ref{sec:firtez}).\label{table:nodes}}
\begin{center}
\begin{tabular}{ccccc}
$\C^{ij}$ & $i=0$ & $i=2$ & $i=4$ & $i=6$ \\
\hline
$T$ & 8 & 8 & 8 & 8 \\
$B_x$ & 1 & 1 & 4 & 4 \\
$B_y$ & 1 & 1 & 4 & 4 \\
$B_z$ & 1 & 1 & 4 & 4 \\
$v_z$ & 1 & 1 & 4 & 4 \\
\end{tabular}
\end{center}
\end{table}

\subsection{Disambiguation module}
\label{sec:disamb}

Between the inversion of the radiative transfer equation (Sect.~\ref{sec:firtez}; $i$-even) and the
MHS module (Sect.~\ref{sec:mhs}; $i$-odd), there is an intermediate step that refers to
the resolution of the 180$^{\circ}$ ambiguity on the horizontal component of the magnetic field. As already 
mentioned in \cite{borrero2019mhs} (Sect.~5), the inversion of the radiative transfer equation provides 
the horizontal component of the magnetic field $(B_x,B_y)$ with an ambiguity of 180$^{\circ}$ \citep{metcalf1994}. 
This means that, at every point on the solar surface $(x,y),$ we could randomly exchange $(B_x,B_y,B_z)$ with 
$(-B_x,-B_y,B_z)$ and the radiative transfer equation would yield exactly the same solution: 
$\varmathbb{I}^{\rm syn}(x,y,\lambda)$. If the magnetic field thus inferred is fed into the MHS module
(Sect.~\ref{sec:mhs}), we would solve for a completely erroneous force balance as the electric currents
derived from such a magnetic field, ${\bf j} = (4\pi)^{-1} c {\bf \nabla} \times {\bf B}$, would be completely unrealistic.

Therefore, we first must ensure that the aforementioned ambiguity has been resolved. While there are many tools 
available to solve this issue \citep{metcalf2006}, we decided to employ the so-called non-potential 
field calculation method (NPFC) from \cite{manolis2005}. Since the NPFC method works in a two-dimensional plane 
parallel to the solar surface (i.e., fixed $z$), we solved the $180^{\circ}$ ambiguity at the height $z$ that corresponds 
to the middle of the sensitivity region for the magnetic field $z=z(\widetilde{\tau})$, where $\widetilde{\tau}$ 
is defined in Eq.~\ref{eq:sensi} (Appendix \ref{app:sensi}). This is where it makes the most sense to solve the 180$^{\circ}$ ambiguity as it
is the region where the errors in the inference of ${\bf B}$ by Firtez-DZ are the smallest.
Elsewhere, we simply extrapolated the solution from the NPFC method to all other $z$ values.

\subsection{Magneto-hydrostatic module}
\label{sec:mhs}

The MHS module receives the physical parameters from the disambiguation module. This is indicated by the green 
arrow in Fig.~\ref{fig:sketch}. The MHS module is based on the approach presented in \cite{borrero2019mhs}. In that paper, we 
employed the "fishpack" library \citep{fishpack1975} to solve the following equation, which represents the MHS equilibrium 
in the solar atmosphere:

\begin{equation}
\nabla^2 P_{\rm g} = -g \frac{\partial \rho}{\partial z} + \frac{1}{c} \nabla \cdot ({\bf j} 
\times {\bf B}) \;\;.
\label{eq:mhsold}
\end{equation}

In this paper, we employed the magnetic field inferred from the inversion of the radiative 
transfer equation. This is not necessarily consistent with the MHD equations and
contains measurement errors \citep[see e.g.,][]{wiegelmann2010mhs}. Therefore, we solved a modified
version of Eq.~\ref{eq:mhsold}, namely

\begin{equation}
\nabla^2 (\ln P_{\rm g}) = -\frac{g u}{K_b} \frac{\partial}{\partial z}\left(\frac{\mu}{T}\right) - 
\frac{f(\beta)}{c P_{\rm g}}\left[\frac{4\pi \|{\bf j}\|^2}{c}+({\bf j} \times {\bf B}) \cdot {\bf \nabla}(\ln P_{\rm g})\right]
\label{eq:mhsnew}
.\end{equation}

The derivation of this equation is detailed in Appendix~\ref{app:mhsnew_deriv}. Here we only need to mention that
the factor $f(\beta)$ is a function that aims at limiting the effect of the Lorentz force in those regions
of the solar atmosphere where the plasma-$\beta$, defined as $\beta = 8\pi P_{\rm g}/\|{\bf B}\|^2$, drops below a certain
value $\beta^{*}$. We prescribe $f(\beta)$ as:

\begin{eqnarray}
f(\beta) = \left\{\begin{tabular}{cc} $(\beta/\beta^{*})^2$ & if $\beta \le \beta^{*}$ \\
$1$ & if $\beta > \beta^{*}$ \end{tabular}\right.
\label{eq:fbeta}
,\end{eqnarray}

\noindent where we adopt $\beta^{*}=0.5$. Using a first estimation of the gas pressure $P_{\rm g}^{i0}$ ($i$-odd), 
we can solve for the left-hand side of Eq.~\ref{eq:mhsnew} as a Poisson-like equation and obtain a new gas 
pressure, $P_{\rm g}^{i1}$, which is then inserted back into the right-hand side, and the process continues until convergence. 
Each time a new gas pressure is obtained, the conversion between $z$ and the optical depth $\tau_c$ 
changes even if the temperature is kept constant (see Appendix~\ref{app:sensi} and Eq.~\ref{eq:ztau}). 
Convergence is assessed by requiring that the Wilson depression $z_w=z(\tau_c=1)$ does not vary, on 
average over the observed region, by more than half a vertical grid point ($\df z/2$) between two consecutive iterations.
To this end, we defined the following $\chi^2$-merit for the Wilson depression:

\begin{eqnarray}
\begin{split}
\chi^2(z_w^{ij+1},z_w^{ij}) = \frac{1}{LM \df z^2} \sum\limits_{l=1}^L\sum\limits_{m=1}^M & [z^{ij+1}(x_l,y_m,\tau_c=1)- \\ &
z^{ij}(x_l,y_m,\tau_c=1)]^2 \;\;\textrm{with $i$-odd}.
\label{eq:chi2z}
\end{split}
\end{eqnarray}

With the previous conditions, convergence is achieved whenever $\chi^{2}(z_w^{ij+1},z_w^{ij})<1/4$. The iterations 
performed by the MHS module are illustrated in Fig.~\ref{fig:sketch} in blue boxes. We will now assume that 
convergence occurs after $j=q$ iterations, resulting in a gas pressure $P_{\rm g}^{iq}$ with $i$-odd. The resulting 
physical parameters $[\C_{\dagger}^{iq},P_{\rm g}^{iq},\rho^{iq}]$ are then sent back into the Stokes inversion module by Firtez-DZ (Sect.~\ref{sec:firtez}). 
We then increase the $i$-index by one, which thus becomes an even number, and again we reset the $j$-index to zero. This is 
indicated by the blue arrow in Fig.~\ref{fig:sketch}.\\

It is important to note here that the physical parameters $\C_{\dagger}$ that the MHS module sends back to the Firtez-DZ
inversion code (blue arrow in Fig~\ref{fig:sketch}) are not exactly the same as the physical parameters $\C$ that the MHS module receives 
from Firtez-DZ (green arrow in Fig~\ref{fig:sketch}). This occurs because even though $C$ and $\C_{\dagger}$ are the same 
in the $(x,y,z)$ domain, they might differ significantly in the $(x,y,\tau_c)$ scale as the conversion
between $z$ and $\tau_c$ is strongly dependent on the gas pressure and density (see Eq.~\ref{eq:ztau}).\\

Finally, it must be borne in mind that, in order to solve Eq.~\ref{eq:mhsnew}, we need to establish a number of 
boundary conditions for the gas pressure $P_{\rm g}$ on the left-hand side of this equation, as well as for the magnetic field and
temperature on the right-hand side. The boundary conditions employed in this work are detailed in Appendix~\ref{app:bc}.

\subsection{Iterating between Firtez-DZ and the MHS module}
\label{sec:iter}

As mentioned in Sect.~\ref{sec:firtez}, the inversion code Firtez-DZ iteratively determines ($i$-even; see red boxes in Fig.~\ref{fig:sketch}) the 
temperature, $T$, the vertical component of the velocity , $v_z$, and three components of the magnetic field, $B_x$, $B_y$, and $B_z$, in the three-dimensional $(x,y,z)$ 
domain. These physical parameters were referred to as $\C$. The gas pressure $P_{\rm g}$ was initially ($i=0$) determined under 
hydrostatic equilibrium (Eq.~\ref{eq:hydeq}), while the density, $\rho$, is determined by applying the equation of state 
(Eq.~\ref{eq:eos}). All these parameters ($\C$, $P_{\rm g}$, and $\rho$) are then passed through the disambiguation module and onto 
the MHS module so as to determine a more consistent gas pressure through the iterative solution of Eq.~\ref{eq:mhsnew} ($i$-odd; 
see blue boxes in Fig.~\ref{fig:sketch}).\\

At this point, at $i=0,$ or in other words before the MHS module has been applied even once, we calculate the gas pressure and 
density through hydrostatic equilibrium (Eq.~\ref{eq:hydeq}). Because $\rho$ depends on the temperature through Eq.~\ref{eq:eos},
we need to reevaluate Eq.~\ref{eq:hydeq} at every step of the $j$-index iteration (Firtez-DZ). That is why in Fig.~\ref{fig:sketch}, 
after the temperature is modified, $T^{0~j+1}=T^{0j}+\delta T^{0j}$, we go back to Eq.~\ref{eq:hydeq} (see solid red arrow). However, 
after the application of the MHS module ($i \ge 1$), the physical parameters are directly employed to solve the radiative transfer 
equation inside Firtez-DZ (blue arrow in Fig.~\ref{fig:sketch}). In fact, for $i \ge 1$, the gas pressure is never modified 
by Firtez-DZ and is kept to whatever values came from the MHS module (dashed red arrow in Fig.~\ref{fig:sketch}; see also 
Sect.~\ref{sec:firtez}). The density, however, does change inside Firtez-DZ because the temperature is being changed by the 
LM and SVD algorithms (LM$+$SVD box in Fig.~\ref{fig:sketch}).\\

Finally, we note that, as mentioned in Sect.~\ref{sec:mhs}, the physical parameters $\C^{ij}_{\dagger}(z)$ that come out of the 
MHS module and are fed back into the Firtez-DZ are in general different from those inferred from the inversion code. Consequently, 
the output physical parameters from the MHS module, $\C^{iq}$ ($i$-odd), will not necessarily produce the same $\varmathbb{I}^{\rm syn}$ 
as the output physical parameters $\C^{ip}$ ($i$-even) from Firtez-DZ. As indicated by the green boxes in Fig.~\ref{fig:sketch}, 
Firtez-DZ verifies this by  measuring whether the physical parameters from the MHS module, $\C^{iq}$, can still produce a good fit 
to $\varmathbb{I}^{\rm obs}$. If they cannot, Firtez-DZ resumes the inversion while keeping the gas pressure fixed at $P_{\rm g}^{iq}$ 
($i$-odd). On the other hand, if $\C^{iq}$ does indeed produce a good fit to $\varmathbb{I}^{\rm obs}$, we can consider that we have 
achieved convergence in both Firtez-DZ and the MHS module, and we therefore exit the process.

\section{Fits to observed data}
\label{sec:fits}

As mentioned in Sect.~\ref{sec:introduction}, one of the limitations in PUS2010 was that the resulting synthetic Stokes
profiles, ${\varmathbb I}^{\rm syn}(x,y,\lambda)$, did not provide the best possible fit to the observed ones, 
${\varmathbb I}^{\rm obs}(x,y,\lambda)$. This was more a matter of choice rather than a real limitation. The optimization 
process in PUS2010, based on a genetic algorithm, was too time-consuming to allow for further iterations in the 
inversion process. Aside from this, there was nothing preventing those authors from feeding their results in the $z$-scale 
back into the Stokes inversion code in order to continue the $\chi^2$-minimization (Eq.~\ref{eq:chi2i}). This is explicitly taken 
into account in our method, as already explained in Sect.~\ref{sec:iter} and illustrated in Fig.~\ref{fig:sketch}. 
Therefore, our method can be considered as having a similar motivation as those from \citet{tino2017invz} and \citet{loeptien2018zw}, in the sense 
that we aim at providing the best possible fit to the observed Stokes profiles. This is in contrast with PUS2010 and 
\citet{andres2019invz}, where fitting the observations plays a secondary role.\\

\begin{figure*}
\begin{center}
\includegraphics[width=18cm]{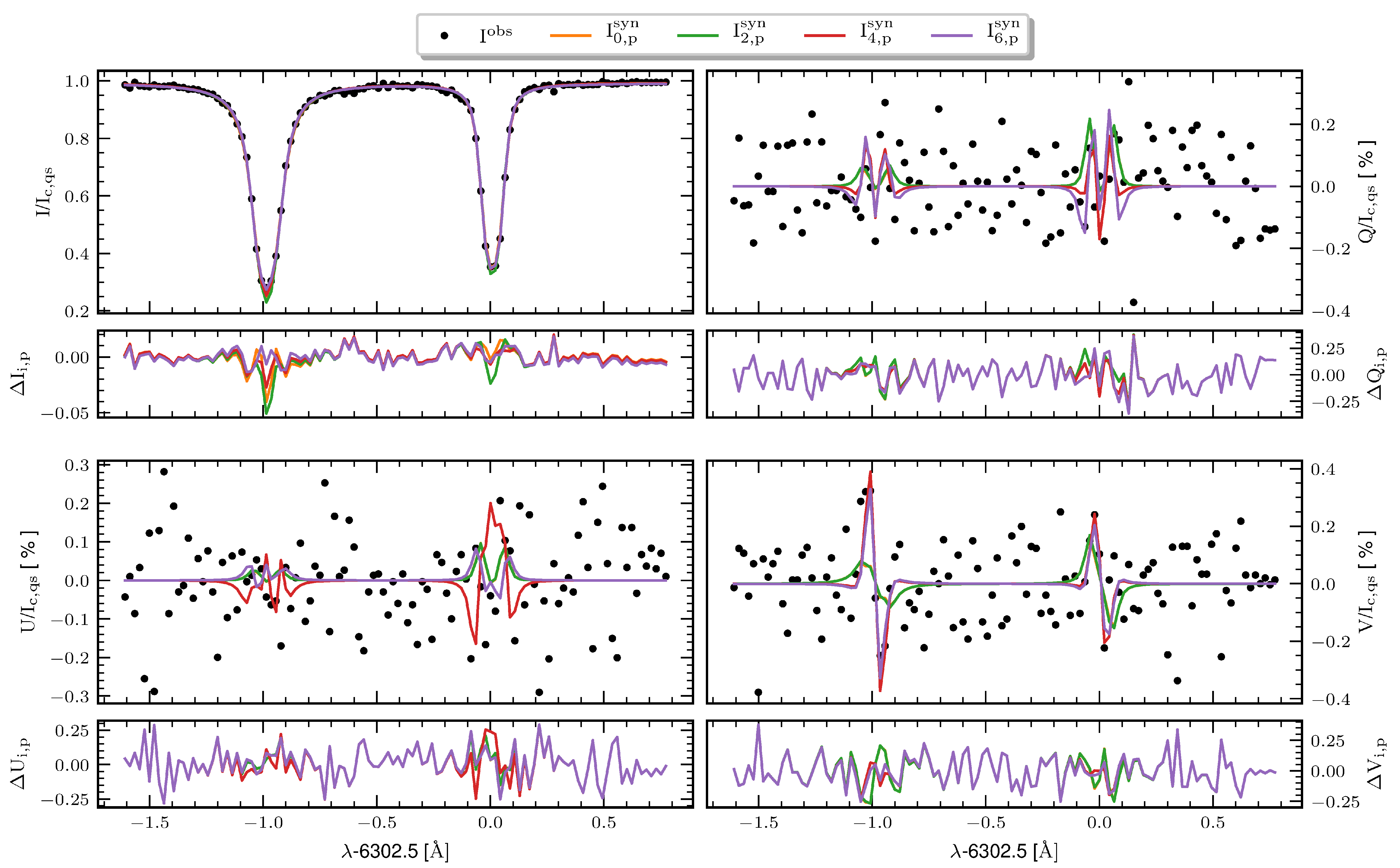}
\caption{Observed (black dots) and best-fit Stokes profiles (solid lines) after the application of the Firtez-DZ
inversion code ($i$-even): $i=0$ (orange), $i=2$ (green), $i=4$ (red), and $i=6$ (purple). The spatial location of these
profiles corresponds to a quiet-Sun pixel (red diamond in the right-hand panel of Fig.~\ref{fig:icmaps}.) The intensity as a function
of wavelength $I(\lambda)$, normalized to the average quiet-Sun continuum intensity $I_{\rm c,qs}$, in the two Fe {\sc I}
lines at 630 nm is presented in the upper-left panel. The linear polarization profiles, $Q(\lambda)$ and $U(\lambda)$,
are displayed in the upper-right and lower-left panels, respectively. Finally, the circular polarization profile, $V(\lambda)$,
is shown in the lower-right panel.\label{fig:fits_qs}}
\end{center}
\end{figure*}

\begin{figure*}
\begin{center}
\includegraphics[width=18cm]{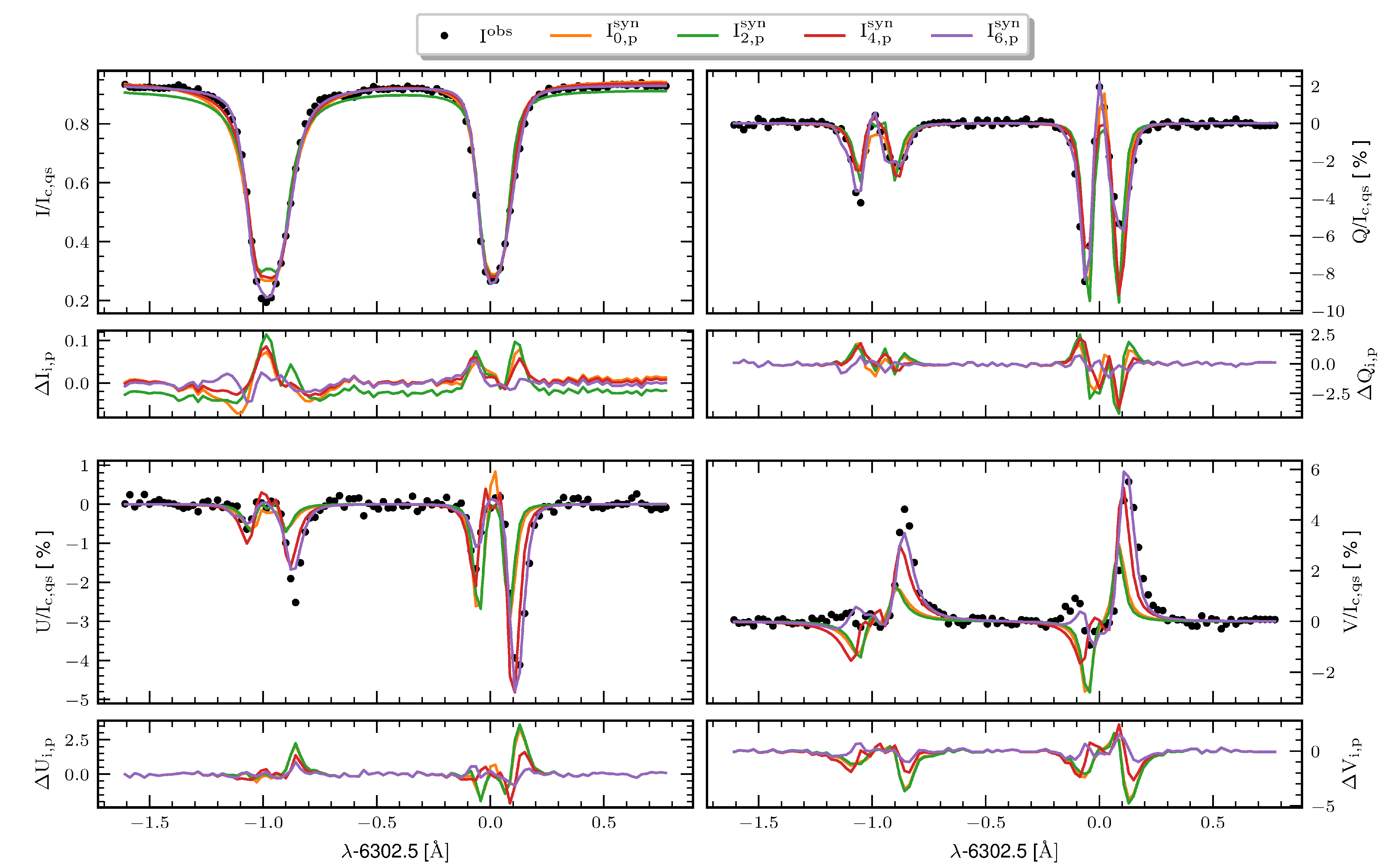}
\caption{Same as Fig.~\ref{fig:fits_qs} but for a pixel located in the penumbra (see the red triangle in the right-hand panel  of 
Fig.~\ref{fig:icmaps}).\label{fig:fits_pen}}
\end{center}
\end{figure*}

\begin{figure*}
\begin{center}
\includegraphics[width=18cm]{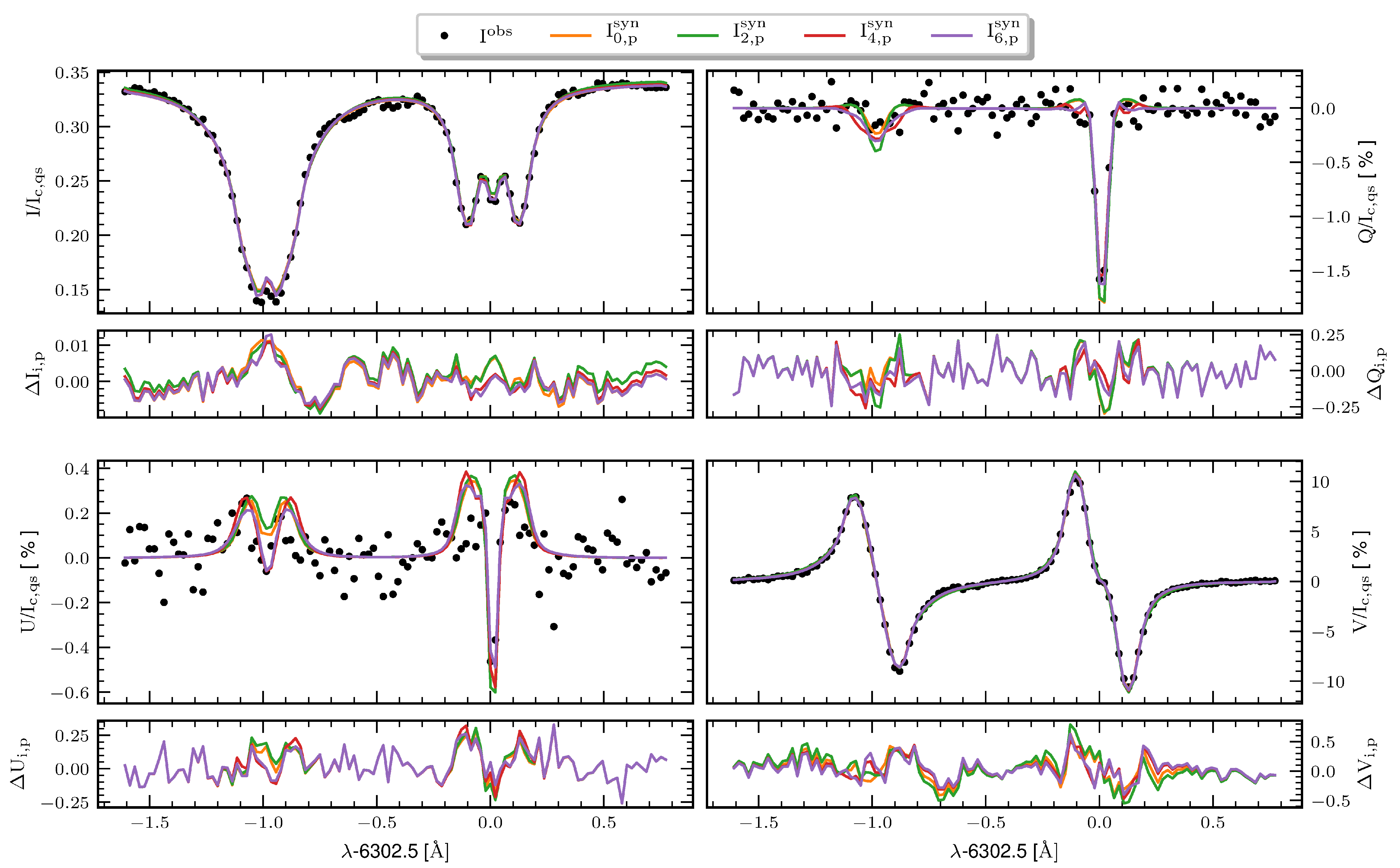}
\caption{Same as Fig.~\ref{fig:fits_qs} but for a pixel located in the umbra (see the red square in the right-hand panel of
Fig.~\ref{fig:icmaps}).\label{fig:fits_umb}}
\end{center}
\end{figure*}

\begin{figure}
\begin{center}
\includegraphics[width=8cm]{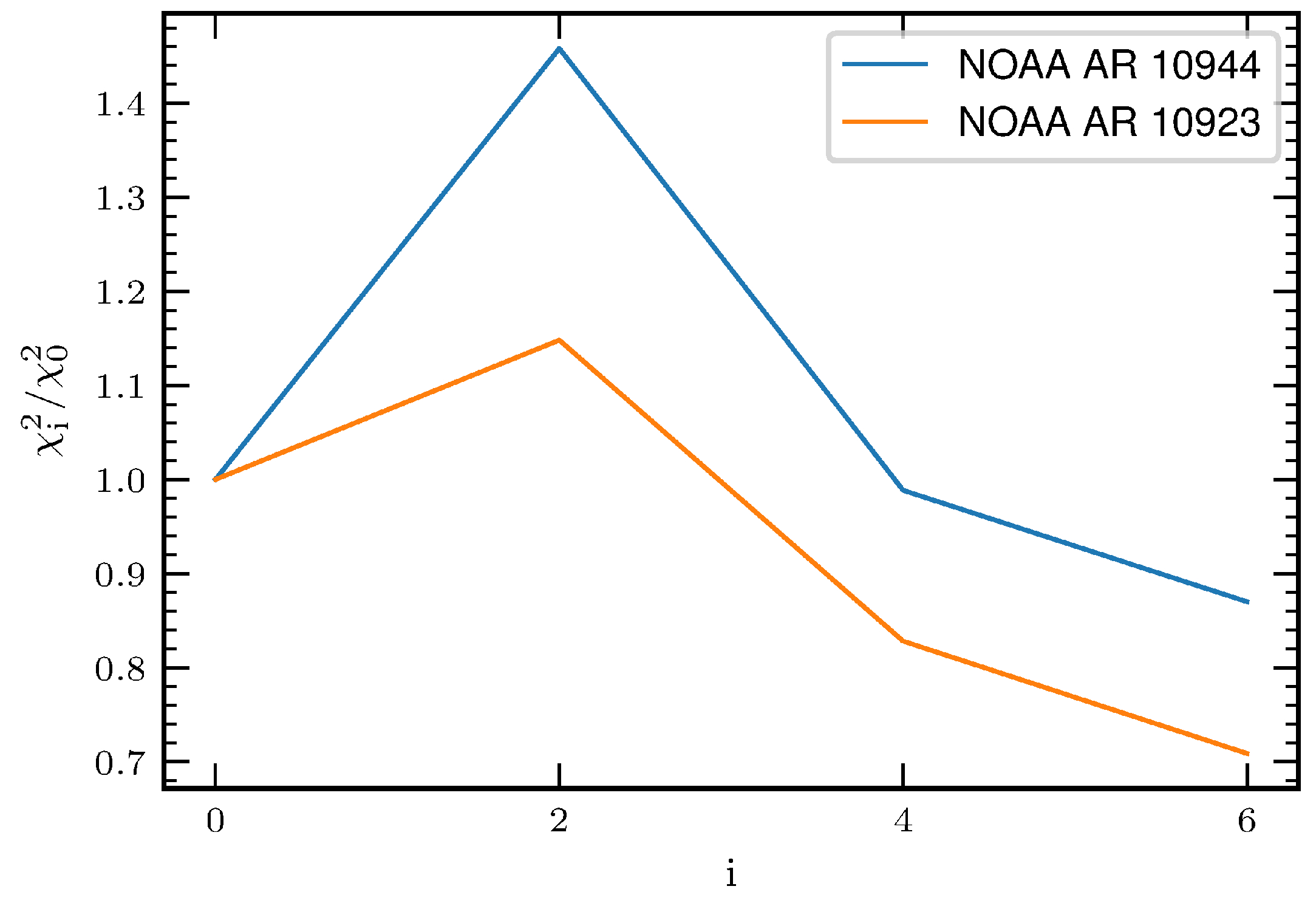}
\caption{Mean value of the $\chi^2$-merit function between the observed ${\varmathbb I}^{\rm obs}(x,y,\lambda)$
and synthetic ${\varmathbb I}^{\rm syn}(x,y,\lambda)$ Stokes profiles (Eq.~\ref{eq:chi2i}) over the entire field-of-view
as a function of the inversion iteration ($i$-even) performed with the Firtez inversion code (Sect.~\ref{sec:firtez}).
Results for NOAA AR 10944 (left-hand panel in Fig.~\ref{fig:icmaps}) are indicated in orange, while results for NOAA AR
10944 (right-hand panel in Fig.~\ref{fig:icmaps}) are shown in blue.\label{fig:chi2iter}}
\end{center}
\end{figure}

To showcase the quality of the fits, we present, in Figs.~\ref{fig:fits_qs}, \ref{fig:fits_pen}, and~\ref{fig:fits_umb}, three 
examples -- in the umbra, penumbra, and quiet Sun, respectively -- of the observed Stokes profiles (black dots) and the best-fit 
profiles (solid colored lines) after $i=0$ (orange), $i=2$ (green), $i=4$ (red), and $i=6$ (purple). These examples provide 
only a qualitative idea about the quality of the fits. A more quantitative picture can be drawn from Fig.~\ref{fig:chi2iter}, 
where we present the mean value of $\chi^2(\varmathbb{I}^{\rm syn}_{ij},\varmathbb{I}^{\rm obs})$ over the entire field-of-view 
for NOAA AR 10944 (blue; right-hand panel in Fig.~\ref{fig:icmaps}) and NOAA AR 10923 (orange; left-hand panel in Fig.~\ref{fig:icmaps}).
As can be seen, $i=6$ yields the best fits of the observed profiles. We note that this is simply a side effect of having 
the largest number of free parameters (see Table~\ref{table:nodes}). This case allows us to fit even well-known asymmetric 
Stokes $V$ profiles found in the sunspot penumbra, as seen in Fig.~\ref{fig:fits_pen} \citep[see also][]{sanchez1992ncp,borrero2006pen}. 
The important thing to consider here is not that the fit improves for larger $i$ values, but rather that it does not get worse. 
The reason is that the pressure and density, and hence 
also the optical-depth scale, are modified after each application of the MHS module ($i$-odd; see Sect.~\ref{sec:mhs}), thus potentially changing the synthetic profiles ${\varmathbb I}^{\rm syn}(x,y,\lambda)$ 
in a way that they no longer provide the best fit to the observed Stokes vector ${\varmathbb I}^{\rm obs}(x,y,\lambda)$ (see 
e.g., Fig.~9 in PUS2010).\\

\section{Inferred physical parameters}
\label{sec:phys}

\begin{figure*}
\begin{center}
\includegraphics[width=14cm]{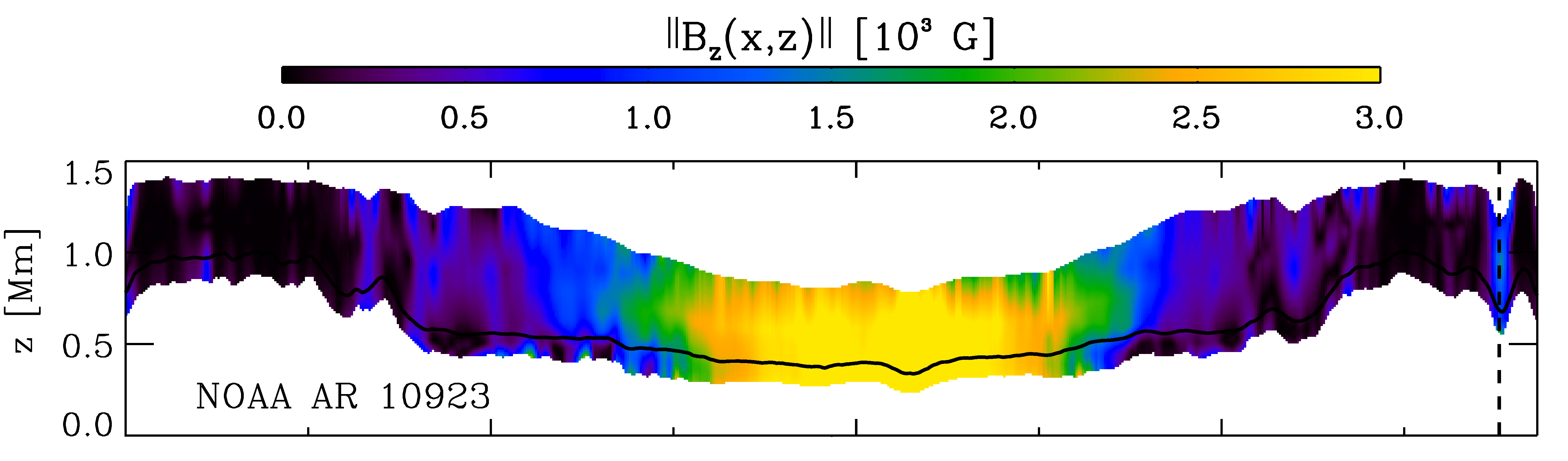} \\
\includegraphics[width=14cm]{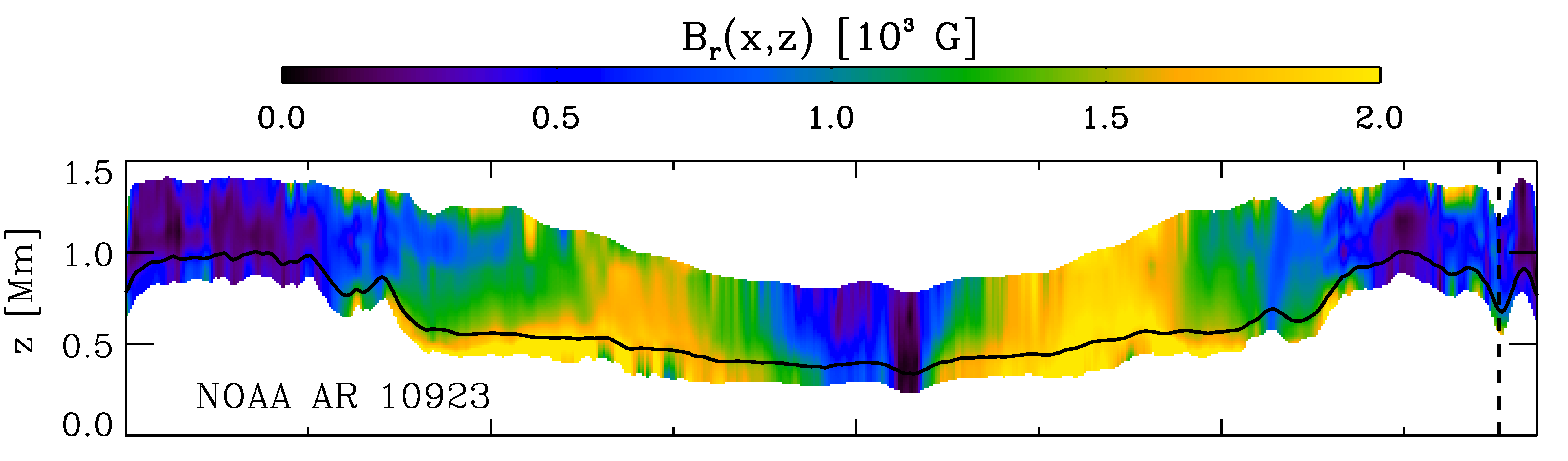} \\
\includegraphics[width=14cm]{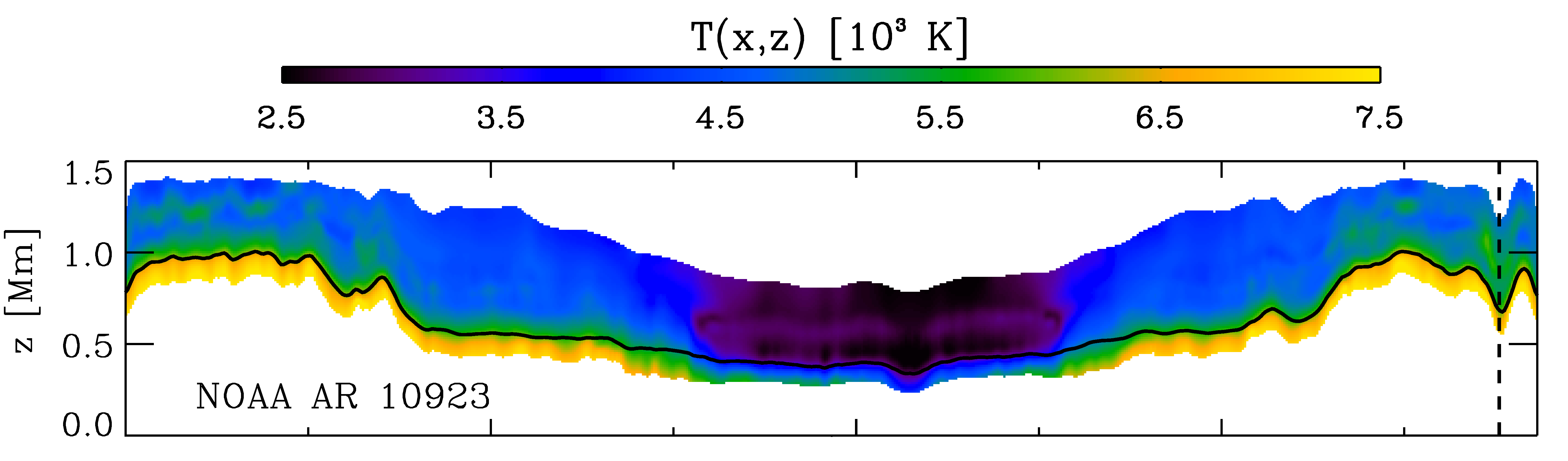} \\
\includegraphics[width=14cm]{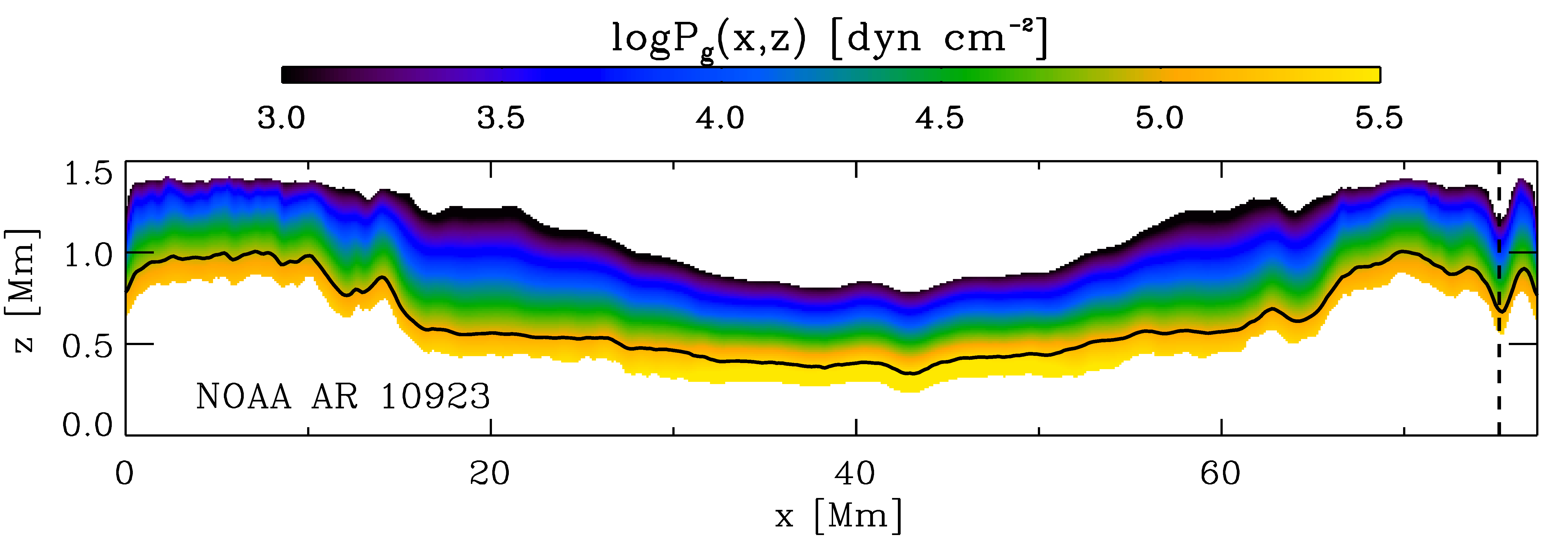}
\caption{Physical parameters for the sunspot NOAA AR 10923 on a vertical slice (XZ-plane) along the blue line in Fig.~\ref{fig:icmaps} 
(left-hand panel). From top to bottom we show: the vertical component of the magnetic field $B_z$, the radial component of the magnetic field $B_r$,
temperature $T$, and the logarithm of the gas pressure $P_{\rm g}$. The solid black line indicates the location of the Wilson
depression ($z(\tau_c=1)$ level). See the text for more details.\label{fig:res_spot1}}
\end{center}
\end{figure*}

\begin{figure*}
\begin{center}
\includegraphics[width=14cm]{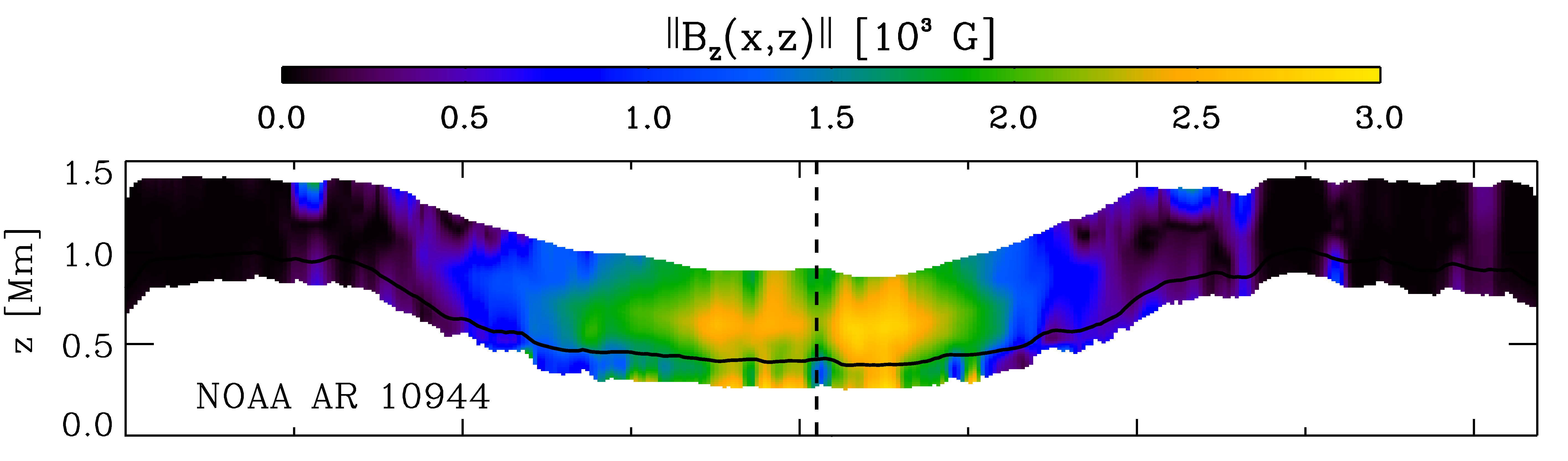} \\
\includegraphics[width=14cm]{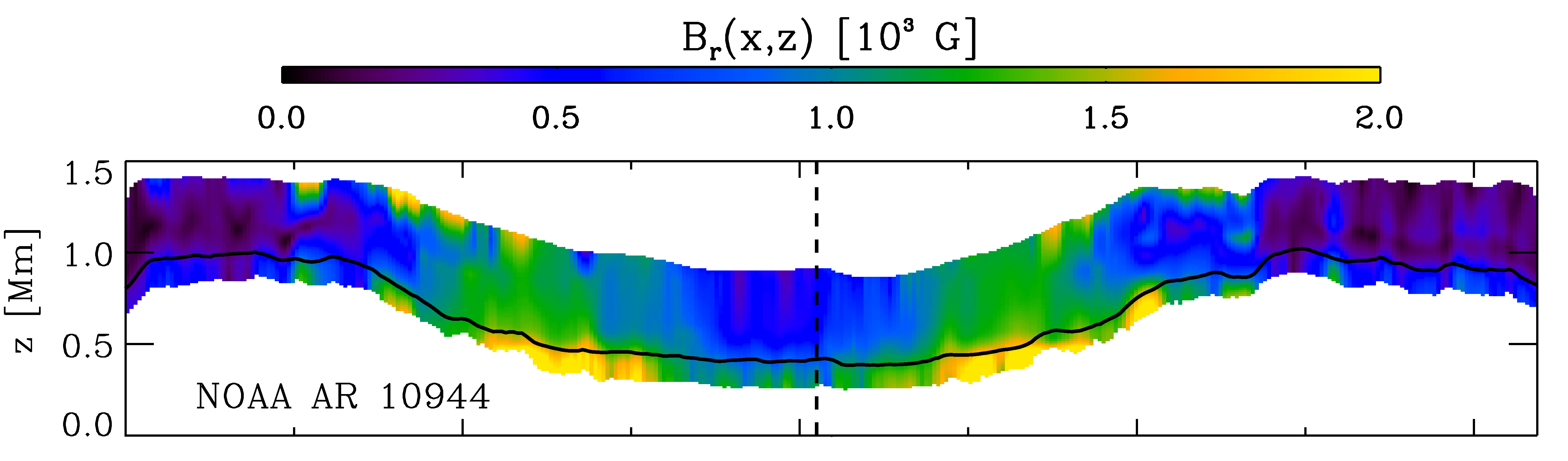} \\
\includegraphics[width=14cm]{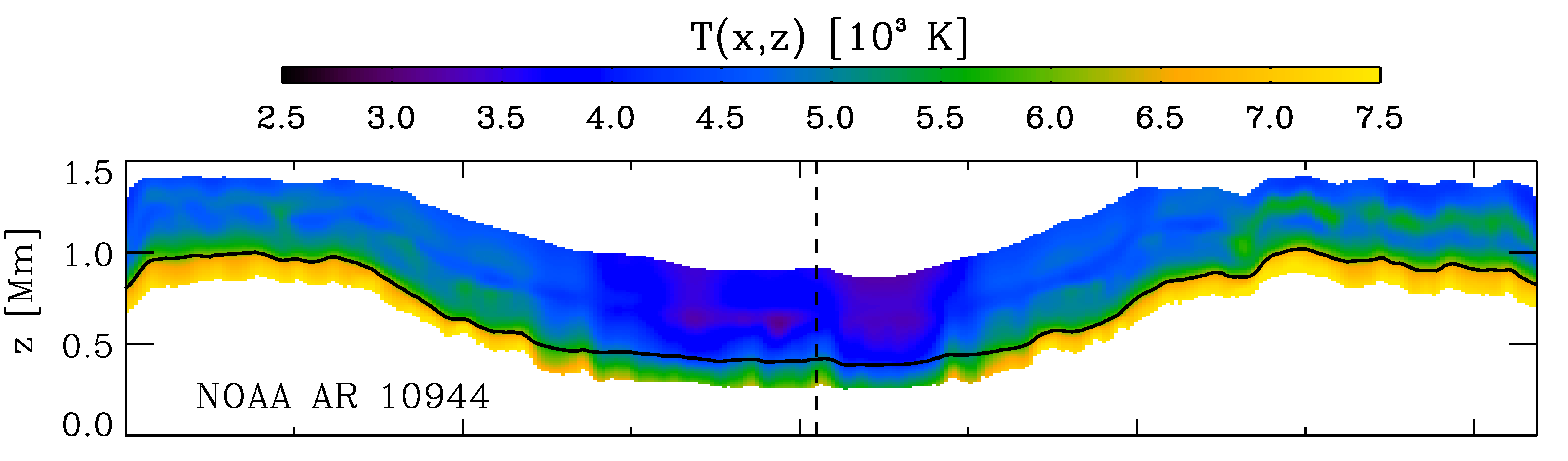} \\
\includegraphics[width=14cm]{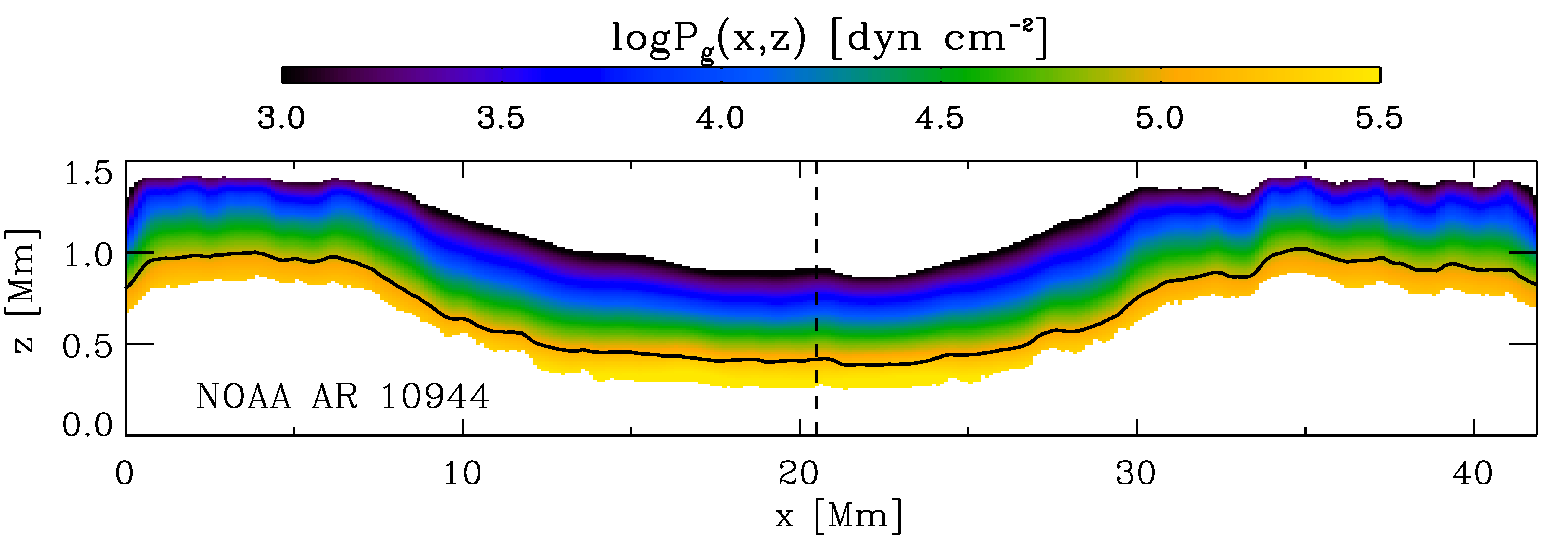}
\caption{Same as Fig.~\ref{fig:res_spot1} but for the sunspot NOAA AR 10944 (right-hand panel in Fig.~\ref{fig:icmaps}).\label{fig:res_spot2}}
\end{center}
\end{figure*}

Next we look at the physical parameters inferred from the combined application of the Firtez-DZ inversion code (Sect.~\ref{sec:firtez})
and the MHS constraints (Sect.~\ref{sec:mhs}). While the physical parameters are retrieved in the $(x,y,z)$ domain, we will not consider 
those regions outside $[z(\tau_a),z(\tau_b)]$, where $\tau_a=10$ and $\tau_b=10^{-4}$. As such, we avoid presenting results in 
atmospheric layers where the errors are large. As explained in Appendix~\ref{app:sensi}, the locations of $z(\tau_a)$ and $z(\tau_b)$ 
depend on the point of the solar surface $(x,y)$ where we look. This can be illustrated by plotting the physical parameters in 
the XZ plane for a fixed value of $y$ (see horizontal blue lines in Fig.~\ref{fig:icmaps}). These physical parameters are presented 
in Figs.~\ref{fig:res_spot1} and ~\ref{fig:res_spot2} for NOAA AR 10923 and 10944, respectively. In these figures, we present the 
absolute value of the vertical component of the magnetic field $\|B_z(x,z)\|$ (first panel), the radial component of the magnetic field 
$B_r(x,z)=[B_x^2(x,z)+B_y^2(x,z)]^{1/2}$ (second panel), the temperature $T(x,z)$ (third panel), and the logarithm of the gas pressure
$\log P_{\rm g}(x,z)$ (fourth panel). We note that NOAA AR 10923 is a negative polarity sunspot ($B_z<0$ in the umbra) but that this is not seen
because we plot only $\|B_z(x,z)\|$. Another important point is that the vertical $z$-scale
and horizontal $x$-scale are not identical in these figures. While the total vertical extension of the box is about 1.5 Mm, it horizontally covers
40-60 Mm (see Sect.~\ref{sec:methodology} and Table~\ref{table:index}). Therefore, for a better visualization, we have stretched
the vertical $z$-scale.\\

In Figs.~\ref{fig:res_spot1} and ~\ref{fig:res_spot2}, the solid black lines indicate the location
of $z(\tau_c=1)$ (i.e., the Wilson depression). In these figures, we can see that, along the selected 
slice of constant $y$ (blue lines in Fig.~\ref{fig:icmaps}), the location of $z(\tau_c=1)$ is about 
$z \approx 1.0$ Mm in the quiet Sun, whereas in the umbra it decreases to about $z \approx 0.4-0.5$ Mm,
yielding a Wilson depression of some 500-600~km. We can also notice many small-scale features. Two
examples of such features are umbral dots and/or light bridges (vertical dashed line in Fig.~\ref{fig:res_spot2}
at $x\approx 21$ Mm), where we see a local enhancement in the temperature $T$ and a local decrease in $B_z$ at 
around $z \approx 0.5$ Mm. This is accompanied by a small increase in the location of the $z(\tau_c=1)$ level. 
Other interesting features are the magnetic field concentrations and magnetic knots outside the sunspot. They are seen,
for instance, at $x \approx 75$ Mm in Fig.~\ref{fig:res_spot1} (see the vertical dashed lines). 
These magnetic knots are characterized by having strong vertical magnetic fields of the same or opposite polarity 
of the sunspot's magnetic field, and they feature a strong dip at the $z(\tau_c=1)$ level.\\

\begin{figure*}
\begin{center}
\includegraphics[width=14cm]{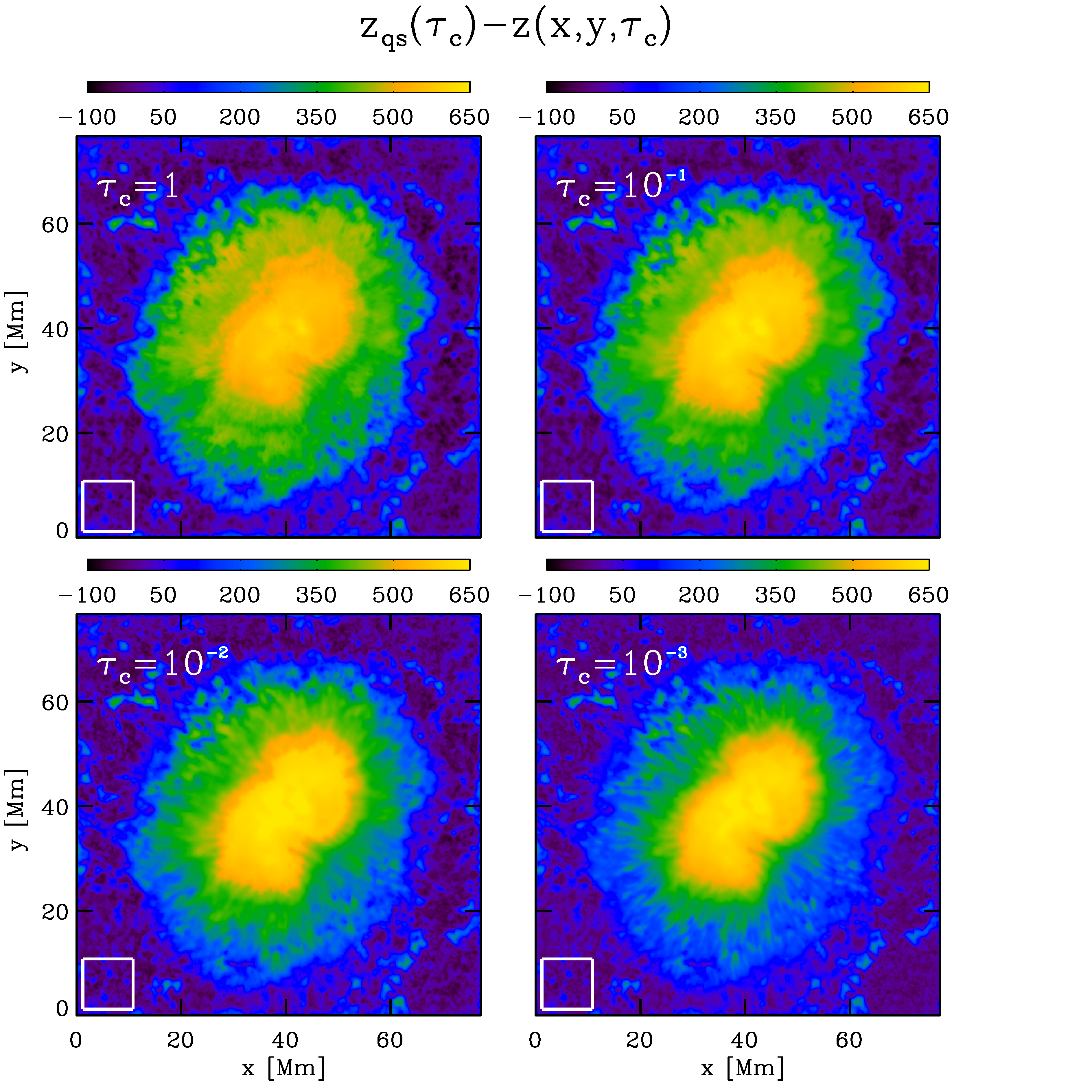}
\caption{Maps of the geometrical height at which different $\tau_c$ levels are reached in NOAA AR 10923: 
$\tau_c=1$ (top left), $\tau_c=10^{-1}$ (top right), $\tau_c=10^{-2}$ (bottom left), and $\tau_c=10^{-3}$ (bottom right). All values 
are given with respect to the geometrical height for that $\tau_c$ level in the quiet Sun, $z_{\rm qs}(\tau_c)$. The quiet-Sun 
value is calculated as the spatial average over the white rectangle.\label{fig:zw_spot1}}
\end{center}
\end{figure*}

\begin{figure*}
\begin{center}
\includegraphics[width=14cm]{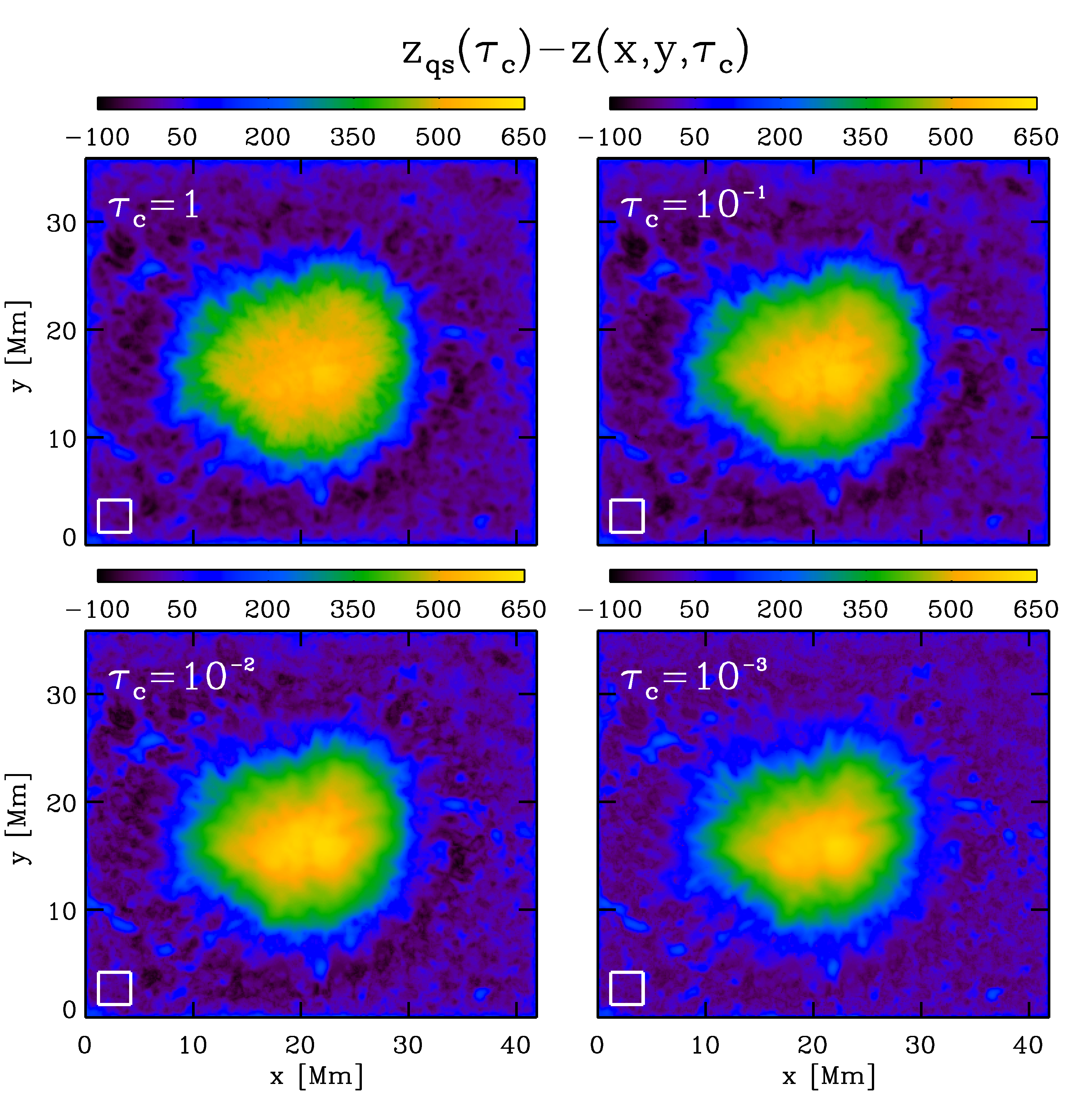}
\caption{Same as Fig.~\ref{fig:zw_spot1} but for NOAA AR 10944.\label{fig:zw_spot2}}
\end{center}
\end{figure*}

In Figs.~\ref{fig:zw_spot1} and ~\ref{fig:zw_spot2}, we show the two-dimensional $(x,y)$ maps of the geometrical
height at which different $\tau_c$ levels are reached in NOAA AR 10923 and 10944, respectively. All values are given 
with respect to the quiet Sun $z_{\rm qs}(\tau_c)$ (see white rectangles in these figures). These
maps correspond to four different realizations of the $z-\tau_c$ conversion (see Eq.~\ref{eq:ztau}) over the entire
observed regions. Again, our method is capable of inferring the small-scale structure of the conversion between
geometrical height $z$ and optical depth $\tau_c$. This is clearly seen around the light bridges in both sunspots
as well as the magnetic knots around them. The mean values of the Wilson depression, $z_{\rm qs}(\tau_c=1)-z(\tau_c=1)$, in 
the umbra obtained with our method are 588 km for NOAA AR 10923 (Fig.~\ref{fig:zw_spot1}; upper-left panel) and 524 
km for NOAA A 10944 (Fig.~\ref{fig:zw_spot2}; upper-left panel). The maximum values around are 630 and 580 km, respectively.\\

It is important to notice that panels for each $\tau_c$ level differ. This is a consequence of our method being capable
of stretching and/or shrinking the $z-\tau_c$ scale between consecutive grid points along the vertical direction
through the changes in temperature, density, and pressure (see Eq.~\ref{eq:ztau}). Other methods, where the $z-\tau$
conversion is obtained by simply shifting, at each $(x,y)$-location, the entire $z$-scale up or down, would yield
exactly the same results at different $\tau_c$ levels in Figs.~\ref{fig:zw_spot1} and ~\ref{fig:zw_spot2}.

\section{Conclusions }
\label{sec:conclusions}

We have presented a new inversion code for the polarized radiative transfer equation that is capable of retrieving
the physical parameters in the solar photosphere in the $(x,y,z)$ domain in a way that is consistent
with the MHS equations, and therefore it takes into account the effects of the
Lorentz force (magnetic tension and pressure) in the force balance. Because of this, our new inversion
code is capable of inferring not only the three components of the magnetic field, the temperature, and the
line-of-sight velocity, but also the gas pressure and density in the solar photosphere. 
The development of this code is inspired, albeit loosely, on the work by \cite{puschmann2010pen}.\\

The inversion code makes use of the Firtez-DZ code and an MHS solver that have been described and tested separately
by \cite{adur2019invz} and \cite{borrero2019mhs}, respectively, employing results from three-dimensional
MHD simulations of sunspots \citep{rempel2012mhd}. In this paper, we combine both approaches
into a single one and test its results with spectropolarimetric observations from the Hinode/SP instrument
in two sunspots located very close to disk center.\\

To put our new approach in context, we will categorize all available methods \citep{carroll2008invz,puschmann2010pen,
tino2017invz, loeptien2018zw,andres2019invz} that also aim at retrieving the physical parameters in the $(x,y,z)$ domain into 
those that: {\bf (a)} can be applied to large regions of the solar surface (i.e., entire sunspots plus their surrounding plage, 
moat, and quiet Sun); {\bf (b)} infer the small-scale structure (i.e., umbral dots, penumbral filaments, light bridges, magnetic 
knots, etc.); and {\bf (c)} fit the observed Stokes vector. All of the aforementioned methods give priority to some features at
the expense of others. For instance, L\"optien et al. (2018) limit the number of Fourier coefficients 
in order to analyze large fields-of-view, thereby limiting their ability to retrieve small-scale structures. Puschmann
et al. (2010) make the opposite sacrifice. The Asensio Ramos \& Diaz Baso (2019) method can deal with both situations but 
does not provide the best possible fit to the observed Stokes vector. Although our method meets the three previous requirements, there
is a very obvious drawback: its speed. Just to give some numbers: the first iteration cycle ($i=0$) took totals of 500 (small spot) 
or 2000 (large spot) combined CPU (central processing unit) hours. Later iterations ($i \ge 2$) needed about half this. Although the actual running time
was significantly reduced by running our inversions in clusters with several hundred nodes, our method is not yet suitable for processing large amounts of data. 
Therefore, which of the available methods is to be preferred depends on the particular use case.\\

Our inversion code also yields, in a natural way, the Wilson depression across the solar surface, not only at $\tau_c=1$
but at all optical depths within the region where the analyzed spectral lines are formed. Our values for the inferred
Wilson depression are compatible, albeit somewhat smaller, by about 50-70 km, with similar studies of the same 
sunspots \citep{loeptien2018zw,andres2019invz,loeptien2020zw}. We note, however, that those studies were carried out
with inversion results that considered the effects of the telescope and instrument point spread function (PSF). Those inversions
usually retrieve sharper variations of the magnetic field along the $(x,y)$-directions, which is likely the reason  our results differ from theirs. 
We will study this particular point in more detail in a future work by implementing the coupled-inversion technique by \cite{vannoort2012decon} into our code,
in order to remove the smearing effects introduced by the instrumental PSF.\\

It is also desirable to check how close to solenoidal the inferred magnetic field ${\bf B}(x,y,z)$ is. This might
imply the implementation of a new approach, within our inversion code, to minimize ${\bf \nabla} \cdot {\bf B}$. Therefore, we 
have decided to leave it for a future study. Such minimization seems to help improve the results of the inferences
in the $(x,y,z)$ domain \citep{puschmann2010pen,loeptien2018zw}. We are not sure, however, how much our method will benefit
from such an implementation as methods that minimize ${\bf \nabla} \cdot {\bf B}$ typically only modify the potential component of the magnetic
field while leaving the non-potential component, and hence the electric currents ${\bf j} \propto {\bf \nabla} \times {\bf B}$, untouched \citep{toth2000divb}.
Consequently, none of those methods would have any effect on the MHS force balance as implemented in our code (Eq.~\ref{eq:mhsnew};
Sect.~\ref{app:mhsnew_deriv}).\\

A corollary of the discussion in the previous paragraph is that, in its current state, our inversion code can be used
to infer realistic electric currents ${\bf j}$ even if the magnetic field is not close to being solenoidal. We foresee
future applications where the full ${\bf j}$-vector, instead of simply its vertical $j_z$-component, is employed to study 
the evolution of magnetic structures on the solar surface that are likely to produce enhanced chromospheric and coronal activity
\citep[see e.g.,][]{solanki2003he,wang2017flare}.

\begin{appendix}
\section{$z-\tau_c$ conversion and sensitivity regions}
\label{app:sensi}

The conversion between the continuum optical depths $\tau_c$ and $z$ depends on the density and continuum opacity $\kappa_{\rm c}$,
which in turn depends on the gas pressure and temperature, as:

\begin{eqnarray}
\df \tau_{\rm c} = -\rho \kappa_{\rm c} (P_{\rm g}, T) \df z \; .
\label{eq:ztau}
\end{eqnarray}

Let us define $z_a=z(\tau_a)$ and $z_b=z(\tau_b)$ as the locations of the optical depths 
$\tau_a$ and $\tau_b$ that cover the sensitivity region of the spectral lines to the physical parameters 
$\C$ as determined by Firtez-DZ. We note that $\tau_a > \tau_b$, whereas $z_a < z_b$ because the optical-depth scale 
and the geometrical scale grow in opposite directions (see Eq.~\ref{eq:ztau}). It is important to bear in mind
that, due to its dependence on the density and opacity, Eq.~\ref{eq:ztau} implies that the locations $z_a$ and $z_b$ 
are different for every $(x,y)$ position.\\

Owing to the fact that different spectral lines are sensitive to different regions in the solar 
atmosphere \citep{basilio94rf}, we adopted $[\tau_a,\tau_b]=[10,10^{-4}]$ in this work. With this, we can define the optical 
depth location that corresponds to the "middle" of the sensitivity region as:

\begin{equation}
\widetilde{\tau} = 10^{\frac{1}{2}[\log\tau_a+\log\tau_b]} \; .
\label{eq:sensi}
\end{equation}

This yields $\widetilde{\tau} \approx 0.0316$. The values of $\tau_a$, $\tau_b$, and $\widetilde{\tau}$ depend, of course,
on the observed spectral lines (see Sect.~\ref{sec:observations}). The more spectral lines that are observed, the larger the sensitivity region becomes.\\

\section{MHS equation}
\label{app:mhsnew_deriv}

Let us start with the momentum equation in ideal MHS \citep[see Eq.~16.23 in][]{kippebook}:

\begin{equation}
\nabla P_{\rm g} = \rho {\bf g} + c^{-1} {\bf j} \times {\bf B} \,
\label{eq:mhsvec}
,\end{equation}

\noindent where $P_g$, $\rho$, and ${g=-g\ez}$  stand for the gas pressure, density, and the Sun's gravity acceleration, respectively. These were introduced in Sect.~\ref{sec:firtez}. The term $c^{-1} {\bf j} \times {\bf B}$ corresponds to the Lorentz
force and can be decomposed into the magnetic pressure and the magnetic tension \citep[Sect.~2.7; Eq.~2.56]{priestbook}. In
\cite{borrero2019mhs}, we took the divergence of this equation to transform it from a system of three first-order partial differential
equations into a single second-order partial differential equation. Here we will proceed along those same lines, but first we
will employ the equation of state (Eq.~\ref{eq:eos}) to substitute the above density, as well as divide the left-hand side and the
right-hand side by the gas pressure:

\begin{equation}
\frac{\nabla P_{\rm g}}{P_g} =  \frac{u}{K_b} \frac{\mu}{T} {\bf g} + \frac{1}{c} \frac{{\bf j} \times {\bf B}}{P_g} \; .
\end{equation}

This equation can be further transformed as follows:

\begin{equation}
\nabla (\ln P_{\rm g}) = -\frac{u g}{K_b} \frac{\mu}{T} \ez + \frac{1}{c} \frac{{\bf j} \times {\bf B}}{P_g} \; .
\end{equation}

Finally, we take the divergence of the equation above, which yields:

\begin{equation}
\nabla^2 (\ln P_{\rm g}) =  -\frac{u g}{K_b} \frac{\partial}{\partial z}\left[\frac{\mu}{T}\right] + 
\frac{1}{c} \ro{\nabla \cdot \left[\frac{{\bf j} \times {\bf B}}{P_g}\right]} \; .
\label{eq:mhsnew_full}
\end{equation}

Unlike Eq.~\ref{eq:mhsold}, solving Eq.~\ref{eq:mhsnew_full} will always yield $P_g>0$. While this was not critical when
employing physical parameters resulting from MHD simulations \citep{borrero2019mhs}, we are now determining the right-hand side using a magnetic field (${\bf B}$) and temperature ($T$) that have been inferred 
from the observations via the inversion of the radiative transfer equation (Sect.~\ref{sec:firtez}). They are therefore 
affected by measurement errors, which become exponentially larger as we consider regions outside the sensitivity region
of the spectral line (see Sect.~\ref{app:sensi}). Consequently, when dealing with actual observations, Eq.~\ref{eq:mhsnew_full}
is highly preferable.\\

We will now focus our attention on the second term on the right-hand side of Eq.~\ref{eq:mhsnew_full} (highlighted in red) 
and expand the divergence operator as:

\begin{equation}
\nabla \cdot \left[\frac{{\bf j} \times {\bf B}}{P_g}\right] = \frac{1}{P_g} [\ro{{\bf \nabla} \cdot ({\bf j}\times{\bf B})} - 
({\bf j}\times{\bf B}) \cdot \nabla (\ln P_g)] \; ,
\label{eq:divlorentzpre}
\end{equation}

\noindent where the first term on the right-hand side (again highlighted in red) of the above equation can be further expanded, 
employing basic vector identities, as:

\begin{equation}
{\bf \nabla} \cdot ({\bf j}\times{\bf B}) = -\frac{c}{4\pi} [(\nabla^2 {\bf B}){\bf B}+\|{\bf \nabla} \times {\bf B}\|^2]
\label{eq:divlorentz}
.\end{equation}

We can see here that the first term on the right-hand side of Equation~\ref{eq:divlorentz} involves second-order spatial derivatives
of the magnetic field, whereas the second term on the right-hand side involves the square of first-order derivatives. Unlike
MHD simulations, where grid sizes are typically on the order of a few kilometers, observational grid sizes are much larger (see 
Sect.~\ref{sec:observations} and Table~\ref{table:index}) and therefore second-order derivatives will be much more inaccurate than
first-order ones. For this reason, we decided to neglect the first term on the right-hand side of 
Equation~\ref{eq:divlorentz} ($\nabla^2 {\bf B}$) and retain only the second term ($\|{\bf \nabla} \times {\bf B}\|^2$).
In the future, it might be possible to include the neglected term as new observing facilities, such as the Daniel K. Inouye
Solar Telescope \citep[DKIST;][]{rimmele2020dkist} and the European Solar Telescope \citep[EST;][]{jan2019est}, will provide spectropolarimetric observations, also with a spatial resolution of a few kilometers. 
Once we insert the simplified Eq.~\ref{eq:divlorentz} into Eq.~\ref{eq:divlorentzpre} and into Eq.~\ref{eq:mhsnew_full} we obtain:

\begin{equation}
\nabla^2 (\ln P_{\rm g}) =  -\frac{u g}{K_b} \frac{\partial}{\partial z}\left[\frac{u}{T}\right]
-\frac{1}{c P_g} \left[\frac{4\pi\|{\bf j}\|^2}{c}+({\bf j}\times{\bf B}) \cdot \nabla (\ln P_g)\right]
\label{eq:mhsnew_simplified}
.\end{equation}

Next we consider that, as we approach the highest layers of the solar photosphere (i.e., close to the temperature minimum),
the density and gas pressure are so low that the Lorentz force term dominates the force balance (Eq.~\ref{eq:mhsvec}). At 
this point, large velocities also usually appear (oftentimes supersonic and super-Alfvenic) so that the advection term 
($\rho ({\bf v}\cdot\nabla)\cdot{\bf v}$) starts to play an important role. Unfortunately, the velocity term is not included 
in our force balance (Eqs.~\ref{eq:mhsold},~\ref{eq:mhsnew},~\ref{eq:mhsvec}) simply because we do not have access, via 
spectropolarimetry, to the horizontal components of the velocity. Until such time that we implement a new method to determine 
$v_x$ and $v_y$ \citep[see e.g.,][]{andres2017vxvy}, we will take a pragmatic approach and consider that the advection term partially 
compensates for the Lorentz force term as we approach regions with very low plasma-$\beta$. To mimic this effect, we introduced a 
scaling function $f(\beta)$ that reduces the effect of the Lorentz force in regions where $\beta \ge 0.5$ (see Eq.~\ref{eq:fbeta}). 
Our approach is justified by the fact that the advection term partially compensates for the 
Lorentz force term in the high photosphere in MHD simulations \citep{rempel2012mhd}, bringing the force balance close to hydrostatic equilibrium. 

\section{Boundary conditions}
\label{app:bc}

In the following, we describe the boundary conditions employed to solve Eq.~\ref{eq:mhsnew}. The need for these boundary conditions
was mentioned in the last paragraph of Sect.~\ref{sec:mhs}.

\subsubsection{$P_{\rm g}$ boundary conditions: Non-axially symmetric sunspots}
\label{subapp:pbc}

The boundary conditions for the gas pressure apply to the left-hand side of Eq.~\ref{eq:mhsnew} and must
be known for all six sides of the three-dimensional volume. These sides are characterized by $x_1=y_1=z_1=0$ 
and by $x_L=L\df x$, $y_M=M\df y$, $z_N=N\df z$ (see Table~\ref{table:index}). In this paper, we consider only 
Dirichlet boundary conditions. In \cite[][see Eq.~9]{borrero2019mhs}, we employed axially symmetric boundary 
conditions for $P_{\rm g}$.  In this paper, we continue using the same values for the side boundaries: 
$P_{\rm}(x_{1},y,z)$, $P_{\rm}(x_{L},y,z)$, $P_{\rm}(x,y_{1},z)$, and $P_{\rm}(x,y_{M},z)$. These 
are adequate as long as the analyzed sunspot is fully surrounded by quiet Sun on all four sides. This
is indeed our case (see Sect.~\ref{sec:observations}). In the $z$-direction, we adopted a different approach that does not assume 
axial symmetry. This is important because, more often than not, sunspots have elliptical shapes, 
contain umbral dots or light bridges, are surrounded by plage or pores, the penumbra is unevenly 
developed, etc. \citep{rolf2010pen,rolf2016lb}. To account for this possibility, we instead employed 
the following {\bf empirical} boundary conditions at $z=z_{1}$ and $z=z_{N}$:

\begin{eqnarray}
\log P_{\rm g}^{ij}(x,y,z_{1}) & = & 6.19 - 4.57\times 10^{-5} \| B_z^{ij}(x,y,\widetilde{\tau})\| \;\; \nonumber \\
\log P_{\rm g}^{ij}(x,y,z_{N}) & = & 2.44 - 9.55\times 10^{-4} \| B_z^{ij}(x,y,\widetilde{\tau})\| \;\;,
\label{eq:pbc}
\end{eqnarray}

\noindent where $\|B_z^{ij}(x,y,\widetilde{\tau})\|$ refers to the modulus of the vertical component of the magnetic field
at an optical depth corresponding to the middle of the sensitivity region $\widetilde{\tau}$ (Eq.~\ref{eq:sensi}). To get 
an idea about the values that Eq.~\ref{eq:pbc} yields, we can consider that, in the quiet Sun, 
$\| B_z(\widetilde{\tau}) \| \approx 0$ Gauss, thus resulting in $P_{\rm g}(z_{1}) \approx 250$ dyn~cm$^{-2}$ 
and $P_{\rm g}(z_{N}) \approx 1.55\times 10^{6}$ dyn~cm$^{-2}$. On the other hand, taking a value of 
$\| B_z(\widetilde{\tau}) \| \approx 4000$ Gauss for a strong umbra, we obtain $P_{\rm g}(z_{1}) \approx 0.04$ 
dyn~cm$^{-2}$ and $P_{\rm g}(z_{N}) \approx 1.02\times 10^{6}$ dyn~cm$^{-2}$. These values are in qualitative agreement
with the results from three-dimensional MHD simulations of sunspots \citep{rempel2012mhd}.\\

The purpose of these boundary conditions is to speed up the convergence of the MHS module (Sect.~\ref{sec:mhs}). 
Using significantly different boundary conditions results in very similar results to those presented in Sect.~\ref{sec:phys}.
We have tested that this is the case by running the MHS module with $P_{\rm g}(z_1)=1.25\times 10^6$ dyn cm$^{-2}$ 
and $P_{\rm g}(z_N)=2.5$ dyn cm$^{-2}$, which are the same at every $(x,y)$, over the lowermost $z=z_1$ and uppermost $z=z_N$ planes. These
results are in agreement with \cite{borrero2019mhs}, where the role of the boundary conditions was studied in more detail.

\subsubsection{$T$ and ${\bf B}$ outside the sensitivity regions}
\label{subapp:tbbc}

At the beginning of Section~\ref{sec:firtez}, we introduced the physical parameters $\C^{ij}=[T, B_x, B_y, B_z]$.
In principle, we could use the physical parameters $\C^{ij}$ inferred from the inversion to solve Eq.~\ref{eq:mhsnew}.
However, the inversion retrieves very unreliable values outside the sensitivity region $[z_a,z_b]$
(see Sect.~\ref{app:sensi}), and therefore we will change the physical parameters outside this region to more meaningful values. 
This will not interfere with our ability to fit the observed Stokes vector because the spectral lines are not sensitive to 
whatever happens outside $[z_a,z_b]$. Consequently, we do not directly employ  $\C^{ij}$ on the right-hand side of Eq.~\ref{eq:mhsnew} 
but rather $\C^{ij}_{\dagger}$, which is constructed from the previous as follows:

\begin{eqnarray}
\C^{ij}_{\dagger}(z) = \left\{\begin{tabular}{cc} $\C(z_1)+\frac{\C^{ij}(z_a)-\C(z_1)}{z_a-z_1}(z-z_1)$ 
& if $z < z_a$ \\ $\C^{ij}(z)$ & if $z \in [z_a,z_b]$ \\
$\C(z_N)+\frac{\C^{ij}(z_b)-\C(z_N)}{z_b-z_N}(z-z_N)$ & if $z > z_b$ \end{tabular}\right.
\label{eq:zdep}
.\end{eqnarray}

Here we see that  $\C_\dagger=\C$ inside the sensitivity region, and therefore we kept the physical parameters
as determined by the Firtez-DZ inversion code (Sect.~\ref{sec:firtez}). Outside the sensitivity region, we 
performed a linear interpolation between $z_1$ and $z_a$
as well as between $z_b$ and $z_N$. Since the values at $z_a$ and $z_b$ are reliable and are provided by the inversion,
all we need to do is establish the values at the boundaries $z_1$ and $z_N$; then, by virtue of the linear 
interpolation in Eq.~\ref{eq:zdep}, we can determine the physical parameters everywhere outside the sensitivity region.
We note that Eq.\ref{eq:zdep} must be applied separately for each $(x,y)$ grid point on the horizontal plane
because $z_a$ and $z_b$ change horizontally (see Sect.~\ref{app:sensi}). Also, it is important to bear in
mind that Eq.~\ref{eq:zdep} must be applied after every $j$-iteration of the solution of Eq.~\ref{eq:mhsnew} (see Sect.~\ref{sec:mhs}) 
as well because $P_g^{ij}$ and $\rho^{ij}$ change with each $j$-iteration and hence so do the locations where $z=z(\tau_a)$
and $z=z(\tau_b)$ (see Eq.~\ref{eq:ztau}).\\

For the temperature at the uppermost boundary, we simply say that $T(z_N)=T(z_b)$, and therefore
the temperature for $z > z_b$ is always constant and equals $T(z_b)$ (i.e., no interpolation needed). 
At the lowermost boundary, we employed a method similar to the one described in \cite{borrero2019mhs} (Sect.~4.2), in which 
we perform azimuthal averages of $T(x,y,z_1)$ as provided by the three-dimensional simulations of sunspots 
\citet{rempel2012mhd} and fit the resulting radial dependence with a fourth-order polynomial. The 
resulting polynomial, as a function of the normalized radial distance $\xi=r/R$ ($R$ is the sunspot radius), is:

\begin{eqnarray}
\begin{split}
\log T(\xi,z_1) & = 3.957 + 0.024 \xi + 0.439 \xi^2 - 0.392 \xi^3 \\ & + 0.094 \xi^4
\end{split}
\label{eq:tbc}
.\end{eqnarray}

Equation~\ref{eq:tbc} yields temperatures of approximately 9000 K and 13500 K at $z_1$ in the center
of the umbra ($\xi=0$) and in the quiet Sun ($\xi=2$), respectively.\\

We will now focus on the horizontal components of the magnetic field. At $z=z_N$ and $z=z_{N-1}$, we consider that
they vanish, whereas at $z=z_1$ we take them to be the same as those in the middle of the sensitivity region
(this last condition also applies to the vertical component of the magnetic field):

\begin{eqnarray}
B^{ij}_x(z_N) & = B^{ij}_x(z_{N-1}) = 0 \notag \\
B^{ij}_y(z_N) & = B^{ij}_y(z_{N-1}) = 0 \notag \\
B^{ij}_x(z_1) & = B^{ij}_x(z[\widetilde{\tau}])\label{eq:bxbybc} \\
B^{ij}_y(z_1) & = B^{ij}_x(z[\widetilde{\tau}]) \notag \\
B^{ij}_z(z_1) & = B^{ij}_z(z[\widetilde{\tau}]). \notag
\end{eqnarray}

The last boundary condition we need is that of the vertical component of the magnetic field at the uppermost 
boundary, $B^{ij}_z(z_N)$. To find it, we first write the radial component of the momentum equation in cylindrical 
coordinates at the uppermost $z$-plane, which, once we apply the boundary conditions for the $B_x$ and $B_y$ components
of the magnetic field given by Eq.~\ref{eq:bxbybc}, simplifies into:

\begin{equation}
\frac{\partial}{\partial r}\left(P_{\rm g}+\frac{B_z^2}{8\pi}\right) = 0 \;.
\label{eq:momr}
\end{equation}

\noindent Using this, we can readily determine the boundary condition for the vertical component of the magnetic field
at $z=z_N$:

\begin{equation}
B^{ij}_z(x,y,z_N) = \sqrt{8\pi[P^{ij}_{\rm g,qs}(z_N)-P^{ij}_{\rm g}(x,y,z_N)]} \;,
\label{eq:bzbc}
\end{equation}

\noindent where the values of the gas pressure at $z_N$ can be obtained from Eq.~\ref{eq:pbc} by inserting
the values of the $z$-component of the magnetic field in the middle of the sensitivity region: $B^{ij}_z(x,y,\widetilde{\tau})$. 
The quiet-Sun values $P^{ij}_{\rm g,qs}$ are obtained by setting the magnetic field in Eq.~\ref{eq:pbc} to zero.\\

\end{appendix}

\begin{acknowledgements}
This work has received funding from the Deutsche Forschungsgemeinschaft (DFG project number 321818926) and from the 
European Research Council (ERC) under the European Union's Horizon 2020 research and innovation programme (SUNMAG, 
grant agreement 759548). JMB acknowledges travel support from the Spanish Ministry of Economy and Competitiveness 
(MINECO) under the 2015 Severo Ochoa Program MINECO SEV-2015-0548 and from the SOLARNET project that has received 
funding from the European Union’s Horizon 2020 research and innovation programme under grant agreement no 824135.
The Institute for Solar Physics is supported by a grant for research infrastructures of national importance from the 
Swedish Research Council (registration number 2017-00625). This research has made use of NASA's Astrophysics Data System. 
Hinode is a Japanese mission developed and launched by ISAS/JAXA, collaborating with NAOJ as a domestic partner, NASA 
and STFC (UK) as international partners. Scientific operation of the Hinode mission is conducted by the Hinode science 
team organized at ISAS/JAXA. This team mainly consists of scientists from institutes in the partner countries. Support 
for the post-launch operation is provided by JAXA and NAOJ (Japan), STFC (U.K.), NASA, ESA, and NSC (Norway)
\end{acknowledgements}

\bibliographystyle{aa}
\bibliography{ms}

\begin{thebibliography}{50}
\expandafter\ifx\csname natexlab\endcsname\relax\def\natexlab#1{#1}\fi

\bibitem[{{Asensio Ramos} \& {D{\'\i}az Baso}(2019)}]{andres2019invz}
{Asensio Ramos}, A. \& {D{\'\i}az Baso}, C.~J. 2019, \aap, 626, A102

\bibitem[{{Asensio Ramos} {et~al.}(2017){Asensio Ramos}, {Requerey}, \&
  {Vitas}}]{andres2017vxvy}
{Asensio Ramos}, A., {Requerey}, I.~S., \& {Vitas}, N. 2017, \aap, 604, A11

\bibitem[{{Bellot Rubio}(2006)}]{luis2006review}
{Bellot Rubio}, L.~R. 2006, in Astronomical Society of the Pacific Conference
  Series, Vol. 358, Solar Polarization 4, ed. R.~{Casini} \& B.~W. {Lites}, 107

\bibitem[{{Borrero} {et~al.}(2014){Borrero}, {Lites}, {Lagg}, {Rezaei}, \&
  {Rempel}}]{borrero2014milne}
{Borrero}, J.~M., {Lites}, B.~W., {Lagg}, A., {Rezaei}, R., \& {Rempel}, M.
  2014, \aap, 572, A54

\bibitem[{{Borrero} {et~al.}(2019){Borrero}, {Pastor Yabar}, {Rempel}, \& {Ruiz
  Cobo}}]{borrero2019mhs}
{Borrero}, J.~M., {Pastor Yabar}, A., {Rempel}, M., \& {Ruiz Cobo}, B. 2019,
  \aap, 632, A111

\bibitem[{{Borrero} {et~al.}(2006){Borrero}, {Solanki}, {Lagg},
  {Socas-Navarro}, \& {Lites}}]{borrero2006pen}
{Borrero}, J.~M., {Solanki}, S.~K., {Lagg}, A., {Socas-Navarro}, H., \&
  {Lites}, B. 2006, \aap, 450, 383

\bibitem[{{Carroll} \& {Kopf}(2008)}]{carroll2008invz}
{Carroll}, T.~A. \& {Kopf}, M. 2008, \aap, 481, L37

\bibitem[{{del Toro Iniesta}(2003{\natexlab{a}})}]{jc2003review}
{del Toro Iniesta}, J.~C. 2003{\natexlab{a}}, Astronomische Nachrichten, 324,
  383

\bibitem[{{del Toro Iniesta}(2003{\natexlab{b}})}]{deltoro2003book}
{del Toro Iniesta}, J.~C. 2003{\natexlab{b}}, {Introduction to
  Spectropolarimetry} (Cambridge, UK: Cambridge University Press, April 2003.)

\bibitem[{{del Toro Iniesta} \& {Ruiz Cobo}(2016)}]{jc2016review}
{del Toro Iniesta}, J.~C. \& {Ruiz Cobo}, B. 2016, Living Reviews in Solar
  Physics, 13, 4

\bibitem[{{Georgoulis}(2005)}]{manolis2005}
{Georgoulis}, M.~K. 2005, \apj, 629, L69

\bibitem[{{Golub} \& {Kahan}(1965)}]{golub1965svd}
{Golub}, G. \& {Kahan}, W. 1965, SIAM Journal on Numerical Analysis, 2, 205

\bibitem[{{Ichimoto} {et~al.}(2007){Ichimoto}, {Suematsu}, {Shimizu},
  {Katsukawa}, {Noguchi}, {Nakagiri}, {Miyashita}, {Tsuneta}, {Tarbell},
  {Shine}, {Hoffmann}, {Cruz}, {Lites}, \& {Elmore}}]{ichimoto2007hinode}
{Ichimoto}, K., {Suematsu}, Y., {Shimizu}, T., {et~al.} 2007, in Astronomical
  Society of the Pacific Conference Series, Vol. 369, New Solar Physics with
  Solar-B Mission, ed. K.~{Shibata}, S.~{Nagata}, \& T.~{Sakurai}, 39

\bibitem[{{Jur{\v{c}}{\'a}k} {et~al.}(2019){Jur{\v{c}}{\'a}k}, {Collados},
  {Leenaarts}, {van Noort}, \& {Schlichenmaier}}]{jan2019est}
{Jur{\v{c}}{\'a}k}, J., {Collados}, M., {Leenaarts}, J., {van Noort}, M., \&
  {Schlichenmaier}, R. 2019, Advances in Space Research, 63, 1389

\bibitem[{{Kippenhahn} \& {Moellenhoff}(1975)}]{kippebook}
{Kippenhahn}, R. \& {Moellenhoff}, C. 1975, Mannheim West Germany
  Bibliographisches Institut AG

\bibitem[{{Kosugi} {et~al.}(2007){Kosugi}, {Matsuzaki}, {Sakao}, {Shimizu},
  {Sone}, {Tachikawa}, {Hashimoto}, {Minesugi}, {Ohnishi}, {Yamada}, {Tsuneta},
  {Hara}, {Ichimoto}, {Suematsu}, {Shimojo}, {Watanabe}, {Shimada}, {Davis},
  {Hill}, {Owens}, {Title}, {Culhane}, {Harra}, {Doschek}, \&
  {Golub}}]{kosugi2007hinode}
{Kosugi}, T., {Matsuzaki}, K., {Sakao}, T., {et~al.} 2007, \solphys, 243, 3

\bibitem[{{Landi Degl'Innocenti} \& {Landi
  Degl'Innocenti}(1985)}]{egidio1985rte}
{Landi Degl'Innocenti}, E. \& {Landi Degl'Innocenti}, M. 1985, \solphys, 97,
  239

\bibitem[{{Lites} {et~al.}(2001){Lites}, {Elmore}, \&
  {Streander}}]{lites2001hinode}
{Lites}, B.~W., {Elmore}, D.~F., \& {Streander}, K.~V. 2001, in Astronomical
  Society of the Pacific Conference Series, Vol. 236, Advanced Solar
  Polarimetry -- Theory, Observation, and Instrumentation, ed. M.~{Sigwarth},
  33

\bibitem[{{L{\"o}ptien} {et~al.}(2018){L{\"o}ptien}, {Lagg}, {van Noort}, \&
  {Solanki}}]{loeptien2018zw}
{L{\"o}ptien}, B., {Lagg}, A., {van Noort}, M., \& {Solanki}, S.~K. 2018, \aap,
  619, A42

\bibitem[{{L{\"o}ptien} {et~al.}(2020){L{\"o}ptien}, {Lagg}, {van Noort}, \&
  {Solanki}}]{loeptien2020zw}
{L{\"o}ptien}, B., {Lagg}, A., {van Noort}, M., \& {Solanki}, S.~K. 2020, \aap,
  635, A202

\bibitem[{{Maltby}(1977)}]{maltby1977zw}
{Maltby}, P. 1977, \solphys, 55, 335

\bibitem[{{Martinez Pillet} \& {Vazquez}(1990)}]{valentin1990zw}
{Martinez Pillet}, V. \& {Vazquez}, M. 1990, \apss, 170, 75

\bibitem[{{Martinez Pillet} \& {Vazquez}(1993)}]{valentin1993zw}
{Martinez Pillet}, V. \& {Vazquez}, M. 1993, \aap, 270, 494

\bibitem[{{Mathew} {et~al.}(2004){Mathew}, {Solanki}, {Lagg}, {Collados},
  {Borrero}, \& {Berdyugina}}]{mathew2004zw}
{Mathew}, S.~K., {Solanki}, S.~K., {Lagg}, A., {et~al.} 2004, \aap, 422, 693

\bibitem[{{Metcalf}(1994)}]{metcalf1994}
{Metcalf}, T.~R. 1994, \solphys, 155, 235

\bibitem[{{Metcalf} {et~al.}(2006){Metcalf}, {Leka}, {Barnes}, {Lites},
  {Georgoulis}, {Pevtsov}, {Balasubramaniam}, {Gary}, {Jing}, {Li}, {Liu},
  {Wang}, {Abramenko}, {Yurchyshyn}, \& {Moon}}]{metcalf2006}
{Metcalf}, T.~R., {Leka}, K.~D., {Barnes}, G., {et~al.} 2006, \solphys, 237,
  267

\bibitem[{{Mihalas}(1970)}]{mihalas1970}
{Mihalas}, D. 1970, {Stellar atmospheres}

\bibitem[{{Pastor Yabar} {et~al.}(2019){Pastor Yabar}, {Borrero}, \& {Ruiz
  Cobo}}]{adur2019invz}
{Pastor Yabar}, A., {Borrero}, J.~M., \& {Ruiz Cobo}, B. 2019, \aap, 629, A24

\bibitem[{{Press} {et~al.}(1986){Press}, {Flannery}, \&
  {Teukolsky}}]{press1986num}
{Press}, W.~H., {Flannery}, B.~P., \& {Teukolsky}, S.~A. 1986, {Numerical
  recipes. The art of scientific computing} (Cambridge: University Press, 1986)

\bibitem[{{Priest}(1984)}]{priestbook}
{Priest}, E.~R. 1984, {Solar magneto-hydrodynamics}

\bibitem[{{Puschmann} {et~al.}(2010){Puschmann}, {Ruiz Cobo}, \&
  {Mart{\'{\i}}nez Pillet}}]{puschmann2010pen}
{Puschmann}, K.~G., {Ruiz Cobo}, B., \& {Mart{\'{\i}}nez Pillet}, V. 2010,
  \apj, 720, 1417

\bibitem[{{Rempel}(2012)}]{rempel2012mhd}
{Rempel}, M. 2012, \apj, 750, 62

\bibitem[{{Riethm{\"u}ller} {et~al.}(2017){Riethm{\"u}ller}, {Solanki},
  {Barthol}, {Gand orfer}, {Gizon}, {Hirzberger}, {van Noort}, {Blanco
  Rodr{\'\i}guez}, {Del Toro Iniesta}, {Orozco Su{\'a}rez}, {Schmidt},
  {Mart{\'\i}nez Pillet}, \& {Kn{\"o}lker}}]{tino2017invz}
{Riethm{\"u}ller}, T.~L., {Solanki}, S.~K., {Barthol}, P., {et~al.} 2017,
  \apjs, 229, 16

\bibitem[{{Rimmele} {et~al.}(2020){Rimmele}, {Warner}, {Keil}, {Goode},
  {Kn{\"o}lker}, {Kuhn}, {Rosner}, {McMullin}, {Casini}, {Lin}, {W{\"o}ger},
  {von der L{\"u}he}, {Tritschler}, {Davey}, {de Wijn}, {Elmore}, {Fehlmann},
  {Harrington}, {Jaeggli}, {Rast}, {Schad}, {Schmidt}, {Mathioudakis},
  {Mickey}, {Anan}, {Beck}, {Marshall}, {Jeffers}, {Oschmann}, {Beard},
  {Berst}, {Cowan}, {Craig}, {Cross}, {Cummings}, {Donnelly}, {de Vanssay},
  {Eigenbrot}, {Ferayorni}, {Foster}, {Galapon}, {Gedrites}, {Gonzales},
  {Goodrich}, {Gregory}, {Guzman}, {Guzzo}, {Hegwer}, {Hubbard}, {Hubbard},
  {Johansson}, {Johnson}, {Liang}, {Liang}, {McQuillen}, {Mayer}, {Newman},
  {Onodera}, {Phelps}, {Puentes}, {Richards}, {Rimmele}, {Sekulic}, {Shimko},
  {Simison}, {Smith}, {Starman}, {Sueoka}, {Summers}, {Szabo}, {Szabo},
  {Wampler}, {Williams}, \& {White}}]{rimmele2020dkist}
{Rimmele}, T.~R., {Warner}, M., {Keil}, S.~L., {et~al.} 2020, \solphys, 295,
  172

\bibitem[{{Ruiz Cobo}(2007)}]{basilio2007review}
{Ruiz Cobo}, B. 2007, in Modern solar facilities - advanced solar science, ed.
  F.~{Kneer}, K.~G. {Puschmann}, \& A.~D. {Wittmann}, 287

\bibitem[{{Ruiz Cobo} \& {del Toro Iniesta}(1994)}]{basilio94rf}
{Ruiz Cobo}, B. \& {del Toro Iniesta}, J.~C. 1994, \aap, 283, 129

\bibitem[{{Sanchez Almeida} \& {Lites}(1992)}]{sanchez1992ncp}
{Sanchez Almeida}, J. \& {Lites}, B.~W. 1992, \apj, 398, 359

\bibitem[{{Schlichenmaier} {et~al.}(2010){Schlichenmaier}, {Rezaei}, {Bello
  Gonz{\'a}lez}, \& {Waldmann}}]{rolf2010pen}
{Schlichenmaier}, R., {Rezaei}, R., {Bello Gonz{\'a}lez}, N., \& {Waldmann},
  T.~A. 2010, \aap, 512, L1

\bibitem[{{Schlichenmaier} {et~al.}(2016){Schlichenmaier}, {von der L{\"u}he},
  {Hoch}, {Soltau}, {Berkefeld}, {Schmidt}, {Schmidt}, {Denker}, {Balthasar},
  {Hofmann}, {Strassmeier}, {Staude}, {Feller}, {Lagg}, {Solanki}, {Collados},
  {Sigwarth}, {Volkmer}, {Waldmann}, {Kneer}, {Nicklas}, \&
  {Sobotka}}]{rolf2016lb}
{Schlichenmaier}, R., {von der L{\"u}he}, O., {Hoch}, S., {et~al.} 2016, \aap,
  596, A7

\bibitem[{{Shimizu} {et~al.}(2008){Shimizu}, {Nagata}, {Tsuneta}, {Tarbell},
  {Edwards}, {Shine}, {Hoffmann}, {Thomas}, {Sour}, {Rehse}, {Ito},
  {Kashiwagi}, {Tabata}, {Kodeki}, {Nagase}, {Matsuzaki}, {Kobayashi},
  {Ichimoto}, \& {Suematsu}}]{shimizu2008hinode}
{Shimizu}, T., {Nagata}, S., {Tsuneta}, S., {et~al.} 2008, \solphys, 249, 221

\bibitem[{{Socas-Navarro}(2001)}]{hector2001review}
{Socas-Navarro}, H. 2001, in Astronomical Society of the Pacific Conference
  Series, Vol. 236, Advanced Solar Polarimetry -- Theory, Observation, and
  Instrumentation, ed. M.~{Sigwarth}, 487

\bibitem[{{Solanki} {et~al.}(2003){Solanki}, {Lagg}, {Woch}, {Krupp}, \&
  {Collados}}]{solanki2003he}
{Solanki}, S.~K., {Lagg}, A., {Woch}, J., {Krupp}, N., \& {Collados}, M. 2003,
  \nat, 425, 692

\bibitem[{{Solanki} {et~al.}(1993){Solanki}, {Walther}, \&
  {Livingston}}]{solanki1993zw}
{Solanki}, S.~K., {Walther}, U., \& {Livingston}, W. 1993, in Astronomical
  Society of the Pacific Conference Series, Vol.~46, IAU Colloq. 141: The
  Magnetic and Velocity Fields of Solar Active Regions, ed. H.~{Zirin},
  G.~{Ai}, \& H.~{Wang}, 48

\bibitem[{{Suematsu} {et~al.}(2008){Suematsu}, {Tsuneta}, {Ichimoto},
  {Shimizu}, {Otsubo}, {Katsukawa}, {Nakagiri}, {Noguchi}, {Tamura}, {Kato},
  {Hara}, {Kubo}, {Mikami}, {Saito}, {Matsushita}, {Kawaguchi}, {Nakaoji},
  {Nagae}, {Shimada}, {Takeyama}, \& {Yamamuro}}]{suematsu2008hinode}
{Suematsu}, Y., {Tsuneta}, S., {Ichimoto}, K., {et~al.} 2008, \solphys, 249,
  197

\bibitem[{{Swarztrauber} \& {Sweet}(1975)}]{fishpack1975}
{Swarztrauber}, P. \& {Sweet}, R. 1975, { Efficient FORTRAN Subprograms for the
  Solution of Elliptic Partial Differential Equations}

\bibitem[{{T{\'o}th}(2000)}]{toth2000divb}
{T{\'o}th}, G. 2000, Journal of Computational Physics, 161, 605

\bibitem[{{Tsuneta} {et~al.}(2008){Tsuneta}, {Ichimoto}, {Katsukawa}, {Nagata},
  {Otsubo}, {Shimizu}, {Suematsu}, {Nakagiri}, {Noguchi}, {Tarbell}, {Title},
  {Shine}, {Rosenberg}, {Hoffmann}, {Jurcevich}, {Kushner}, {Levay}, {Lites},
  {Elmore}, {Matsushita}, {Kawaguchi}, {Saito}, {Mikami}, {Hill}, \&
  {Owens}}]{tsuneta2008hinode}
{Tsuneta}, S., {Ichimoto}, K., {Katsukawa}, Y., {et~al.} 2008, \solphys, 249,
  167

\bibitem[{{van Noort}(2012)}]{vannoort2012decon}
{van Noort}, M. 2012, \aap, 548, A5

\bibitem[{{Wang} {et~al.}(2017){Wang}, {Liu}, {Ahn}, {Xu}, {Jing}, {Deng},
  {Huang}, {Liu}, {Kusano}, {Fleishman}, {Gary}, \& {Cao}}]{wang2017flare}
{Wang}, H., {Liu}, C., {Ahn}, K., {et~al.} 2017, Nature Astronomy, 1, 0085

\bibitem[{{Wiegelmann} \& {Inhester}(2010)}]{wiegelmann2010mhs}
{Wiegelmann}, T. \& {Inhester}, B. 2010, \aap, 516, A107

\end{thebibliography}

\end{document}